\documentclass[twocolumn,showpacs,amssymb,nobibnotes,aps,floats,psfig,prb]{revtex4-1}
\usepackage{textcomp,amssymb,graphicx,epsf}
\usepackage{hyperref}
\usepackage{amsbsy}
\usepackage{amsmath}
\usepackage {xcolor}
\usepackage{graphicx,epsfig}
\usepackage{float}
\begin{document}
\title{
Giant magnetoelectric effect in
 pure manganite-manganite 
 heterostructures}
\author{Sanjukta Paul${^1}$} 
\author{Ravindra Pankaj${^1}$}
\author{Sudhakar Yarlagadda${^{1}}$}
\author{ Pinaki Majumdar${^{2}}$}
\author{Peter B. Littlewood${^{3,4}}$}
\affiliation{${^1}$CMP Div., Saha Institute of Nuclear Physics, HBNI,
Kolkata, India}
\affiliation{${^2}$ Harish-Chandra Research Institute, HBNI, Allahabad, India}
\affiliation{${^3}$Argonne National Laboratory, Argonne IL 60439}
\affiliation{${^4}$University of Chicago,James Franck Institute, Chicago IL 60637}
\date{\today}
\begin{abstract}
Obtaining strong magnetoelectric couplings in  bulk materials and heterostructures
is an ongoing challenge. We demonstrate that manganite 
heterostructures of the form  ${\rm (Insulator)/(LaMnO_3)_n/(CaMnO_3)_n/(Insulator)}$ 
show strong multiferroicity in magnetic manganites where 
ferroelectric polarization is realized by charges leaking from  ${\rm LaMnO_3}$
 to  ${\rm CaMnO_3}$ due to repulsion.
Here, an  effective nearest-neighbor electron-electron (electron-hole) repulsion (attraction)
is generated by cooperative electron-phonon
interaction.
Double exchange, when a particle 
virtually hops to its unoccupied neighboring site and back,
produces magnetic polarons
that polarize  antiferromagnetic regions.
Thus a striking giant magnetoelectric effect ensues when
an external electrical field enhances the electron leakage across the interface.
\end{abstract}

\maketitle

\pagebreak

\section{Introduction} 
Complex oxides such as manganites display a rich interplay of various 
orbital, charge, and spin orders when rare-earth dopants are added to the
parent oxide. While significant progress has been made in characterizing
the bulk doped materials, the heterostructures produced from two parent
oxides is an area of active research \cite{ahn,millis0}. In these
heterostructures, the accompanying  quantum confinement, anisotropy,
heterogeneity, and the
enhanced gradients (in magnetic moments, electric potential, orbital polarization, etc.) across
the interface result in novel phenomena that
have no counter part in the bulk samples \cite{pbl1}.
In fact, the challenge is to 
technologically exploit this new physics and develop new useful
devices that can meet future 
demands such as miniaturization,
dissipationless operation/manipulation (such as read/write capability),
energy storage, etc \cite{mannhart0,mannhart01,hwang0,spaldin0,tokura0,dagotto0,ramesh0}. 

There has been a revival in multiferroic research partly due to
improved technology, discovery of new compounds 
(such as ${\rm YMnO_3}$, ${\rm TbMn_2 O_5}$),
need for devices with strong magnetoelectric effect,
etc \cite{tsymbal1}. Majority of the multiferroics studied are bulk materials
where it is not yet clear why the magnetic and electric polarizations
coexist poorly \cite{khomskii1}.
With the advent of improved molecular-beam-epitaxy
technology one can now grow 
oxide heterostructures 
with atomic-layer precision
and explore the
possibility of strong multiferroic phenomena as well as large
interplay between ferroelectricity and magnetic polarization.
Coupling between the charge and
spin degrees of freedom is fascinating
 both from a fundamental viewpoint
 as well as from an applied perspective. Instead of employing currents and
magnetic fields, controlling and manipulating 
 magnetism  with electric fields holds a lot of promise
 as the electric fields are easier
to use in smaller dimensions
and can potentially lower energy
consumption in systems.
There are numerous mechanisms for magnetoelectric effect; reviews for
these can be found in Refs. \onlinecite{khomskii2,wang,spaldin,ramesh1,sdong}.
At the interface of a magnetic oxide and a ferroelectric/dielectric oxide, 
 magnetoelectric effect of electronic origin has
been predicted by some researchers. Upon application
of an external electric field, not only
the magnitude
of moments can be changed \cite{spaldin1,me9}, but in some cases
the very nature of magnetic ordering can be changed \cite{tsymbal2}.

Among various efforts pertaining to oxide heterostructures, there is
considerable interest, both experimentally 
\cite{anand1,anand2,anand3,anand4,schiffer1,kawasaki1,kawasaki2,schlom}
as well as theoretically 
\cite{millis2,millis3,satpathy,dagotto3,dagotto1,perroni},
in understanding  novel aspects of conductivity, magnetism, and orbital order
in pure manganite-manganite
${\rm TMnO_3/DMnO_3}$ heterostructures
where T refers to trivalent rare earth elements La, Pr, Nd, etc.
and D refers to divalent alkaline elements Sr, Ca, etc.
At low temperatures, the bulk ${\rm TMnO_3}$
is  an insulating A-type antiferromagnet (A-AFM); on the other hand,
the bulk ${\rm DMnO_3}$ is an insulating G-type antiferromgnet (G-AFM).
Furthermore, the doped alloy ${\rm T_{1-x} D_x MnO_3}$ 
is an antiferromagnet for $x > 0.5$; whereas for $x < 0.5$,
it is a ferromagnetic insulator (FMI) at smaller values of x (i.e., $0.1 \lesssim x \lesssim 0.2$)
\cite{tokura,cheong,raveau} and is a ferromagnetic metal at higher dopings
in ${\rm La_{1-x} Sr_x MnO_3}$, ${\rm La_{1-x} Ca_x MnO_3}$, ${\rm Pr_{1-x} Sr_x MnO_3}$,
and ${\rm Nd_{1-x} Sr_x MnO_3}$.
{For representative studies of doped manganite-manganite heterostructures, the
reader is referred to Ref. \onlinecite{kalpataru}}.

In spite of considerable efforts towards control of magnetization
through electric fields in multiferroic bulk materials and heterostructures,
obtaining strong magnetoelectric couplings continues to be a challenge. 
Here, in this paper, we predict a novel giant magnetoelectric effect,
not at the interface, but away from it, in 
a pure manganite-manganite heterostructure (see Fig. 1).
Our objective is to present a plausible
  multiferroic phenomenon in manganite heterostructures and point
out the associated unnoticed
striking magnetoelectric effect.
Cooperative electron-phonon 
interaction is shown to be key to understanding both multiferroicity and magnetoelectric
effect in our oxide heterostructure.
Here, we exploit the fact that manganites have
various competing phases  that are close in energy and that by using
an external perturbation (such as an electric or a magnetic field)
the system can be induced to alter its phase.
We show that
there is a charge redistribution (with a net electric dipole moment perpendicular
to the interface) and a concomitant ferromagnetism due to 
the optimization produced by the following two competing effects:
(i) energy cost to produce
holes on the ${\rm LaMnO_3}$ (LMO) side and excess electrons on the ${\rm CaMnO_3}$ (CMO) side; and
(ii) energy gain due to electron-hole attraction (or electron-electron repulsion)
on nearest-neighbor ${\rm Mn}$ sites
induced by electron-phonon interaction.
 The charge polarization is
akin to that of a pn-junction in semiconductors although the
governing equations are different.
The key results of our analysis are as follows:
(i) the interface charge density at the LMO-CMO interface is 0.5 electrons/site.
The LMO-CMO interface is ferromagnetic and persists to be ferromagnetic 
for another layer adjacent to the interface on the LMO side; 
(ii) minority carriers leak across the interface of the heterostructure and
produce ferromagnetic domains due to the  
ferromagnetic coupling (generated by electron-phonon interaction and double-exchange)
between an electron-hole pair on adjacent sites; and
(iii) since ferroelectricity and ferromagnetism
have a common origin [i.e., minority carriers or holes (electrons) on LMO (CMO) side],
there is a striking interplay between these two polarizations;  consequently, when an external electric
field is applied to increase minority carriers, 
 a giant magnetoelectric effect results.

The rest of the paper is organized as follows. In Sec. II we introduce our phenomenological
Hamiltonian (based on cooperative electron-phonon-interaction physics) using which we deduce the charge
distribution in the continuum approximation. We also provide a simple
analytic treatment for the magnetic profile and demonstrate a giant magnetoelectric
effect in a few-layered heterostructure.
Next, in Sec. III, we adopt a more detailed numerical approach and introduce
a Hamiltonian that includes additional kinetic terms, work functions, and discrete lattice 
effects. Here, magnetoelectric effect is studied in
symmetric lattices (involving equal number of LMO and CMO layers)
as well as in asymmetric lattices  using Monte Carlo simulations. 
We close in Sec. IV with our concluding observations.

\section{Analytic treatment}
\label{sec2}
We will begin our treatment of the pure manganite-manganite  heterostructure
by considering a simple analytic picture in this section.
In the next section, we will take recourse to a more detailed numerical approach. 
\subsection{Polaronic Hamiltonian}
In our ${\rm (Insulator)/(LaMnO_3)_n/(CaMnO_3)_n/(Insulator)}$ heterostructure
depicted in Fig. 1, 
{due to charge leaking across the LMO-CMO interface,
we expect different states of the phase diagram of ${\rm La_{1-x}Ca_x MnO_3}$ (LCMO)
at different cross-sections perpendicular to the growth direction.}
Since far from the LMO-CMO
interface the material properties must be similar to those in the bulk,
we expect the ${\rm x=0}$
phase at the Insulator-LMO interface  and 
the  ${\rm x =1}$ phase at the other end involving CMO-Insulator interface.
Considering majority of the LCMO phase diagram (including the end regions near $x=0$ and $x=1$)
is taken up by insulating phases,
since band width is significantly diminished at strong electron-phonon coupling,
and because the  heterostructures are quasi two-dimensional (2D),
we expect that there
is no effective transport in the direction normal to the oxide-oxide interface (i.e., the z-direction).
Then, {
for analyzing the charge distribution normal to the interface},
the starting polaronic Hamiltonian is assumed to comprise of localized electrons and have the following
phenomenological form:
\begin{equation}
H_
{\rm pol}\sim -\sum_{j,\delta}
\left [ \gamma^1_{\rm ep} g^2\omega_0 +\frac{\gamma^2_{\rm ep} t_{j,j+\delta}^2}{g^2\omega_0} \right ]
n_j(1-n_{j+\delta}),
\label{H_pol}
\end{equation}
where the first coefficient $\gamma^1_{\rm ep} g^2 \omega_0$ is due to electron-phonon interaction
and represents nearest-neighbor electron-electron repulsion brought
about by incompatible distortions of nearest-neighbor oxygen cages surrounding occupied ${\rm Mn}$ ions.
The pre-factor $\gamma^1_{\rm ep}$ can depend on the phase -- for instance,
 in the regime of C-type antiferromagnet (C-AFM)
in LCMO, $\gamma^1_{\rm ep}$ is expected to be large because occupancy of
neighboring $d_{z^2}$ orbitals is inhibited
in the z-direction; while in the regime of A-AFM  (corresponding to undoped ${\rm LaMnO_3}$),
$\gamma^1_{\rm ep}$ is expected to be weaker because of
compatible Jahn-Teller distortions on neighboring sites. Here, we will assume for
simplicity that $\gamma^1_{\rm ep}$ is concentration independent
and that $0.1 \le \gamma^1_{\rm ep} \le 1 $. Next,
the coefficient $\gamma^2_{\rm ep} t^2/( g^2 \omega_0)$ results from processes involving hopping to
 nearest-neighbor and back and is present even when we consider the simpler
Holstein model \cite{sdadsy1} or the Hubbard-Holstein model \cite{srsypbl1}.
The pre-factor $\gamma^2_{\rm ep}$  varies between
 $1/2$ (for non-cooperative electron-phonon interaction)
and $1/4$ (since for cooperative  breathing
mode 
in one-dimensional chains $\gamma^2_{\rm ep} =1/3$, which should be more than in C-chains)
\cite{rpsy}.  
Now, even within the two-band picture of  manganites in Ref. \onlinecite{tvr},
the electrons in the localized polaronic band  contribute the term $t^2/g^2 \omega_0$; the 
broad band (due to undistorted states that are orthogonal to the polaronic states) is an upper
band whose band width is reduced due to the 2D nature of the system and does
not overlap with the polaronic band to produce conduction even at carrier concentrations
corresponding to
${\rm 0.2 \lesssim x \lesssim 0.5}$.
Furthermore, although $n_j$ is the total number in both the orbitals at site $j$,
it can only take a maximum value of 1 due to strong on-site electron-electron repulsion
and strong Hund's coupling. 
 Next, to make the above Hamiltonian furthermore relevant for manganites, one needs 
to consider Hund's coupling between core $t_{\rm 2g}$ spins and itinerant $e_{\rm g}$ electrons.
 This leads to invoking the double exchange mechanism for transport.
Then, the hopping term $t_{i,j}$ between sites $i$ and $j$ in Eq. (\ref{H_pol})
 is modified to be  $t_{i,j}\sqrt{0.5[1+({\bf S}
 _i\cdot{\bf S}_j/{\bf S}^2)]} =t_{i,j}\cos(\theta_{ij}/2)$ with ${\bf S}_i$ being 
 the core $t_{\rm 2g}$ spin at site $i$ and $\theta_{ij}$ being the angle between 
 ${\bf S}_i$ and ${\bf S}_j$. The term ${\gamma^2_{\rm ep} t_{j,j+\delta}^2\cos^2(\theta_{ij}/2)}/(g^2\omega_0)$
in Eq. (\ref{H_pol}) produces a strong ferromagnetic coupling between the spins  at site $j$
and  site $j+\delta$ and this dominates over any superexchange coupling between the two spins.
\begin{figure}[h]
\includegraphics[width=2.4in,height=2.2in]{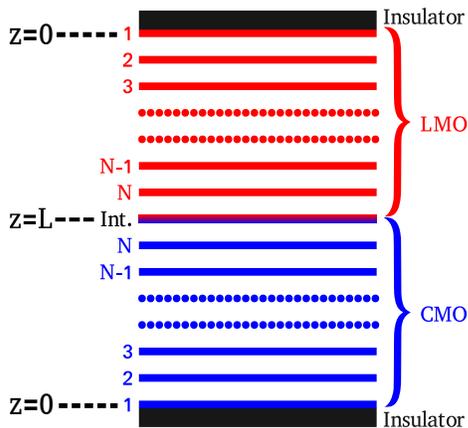}
\caption{(Color online) Schematic showing the symmetric
 ${\rm (Insulator)/(LaMnO_3)_N/(CaMnO_3)_N/(Insulator)}$ heterostructure.
Each of the labeled $N$ layers on both LMO (${\rm LaMnO_3}$) and CMO (${\rm CaMnO_3}$) sides, as
well as in the Interface, comprise of manganese-oxide (MO) layers.
}
\label{fig:het_stut}
\end{figure}

\subsection{Charge Profile}
\label{analytic_charge}
To obtain the charge distribution,
we ignore the effect of superexchange interaction in the starting effective Hamiltonian
in Eq. (\ref{H_pol}) because its energy scale is significantly
smaller than the polaronic energy term $2 g^2 \omega_0$.
For a localized system,
we only need to minimize the interaction energy which is a functional
of the electronic density profile.
The Coulombic interaction energy resulting from the electrons leaking from the LMO side
to the CMO side, is taken into account by ascribing an effective
 charge $+1$ (hole)
to the
LMO
unit cell (centered at the ${\rm Mn}$ site)
that has donated an $e_{\rm g}$ electron from the ${\rm Mn}$ site
and an effective charge $-1$ (electron) to the CMO
unit cell (centered at the ${\rm Mn}$ site)
that has accepted an $e_{\rm g}$ electron at the ${\rm Mn}$ site. 
The Coulombic energy that results is due to the interactions between
these $\pm 1$ effective charges. 
The net positive charge on the LMO side and the net negative charge
on the CMO side will produce a charge polarization (or inversion asymmetry).
 Since the ferroelectric dipole is
expected to be in the 
direction 
perpendicular to the oxide-oxide (LMO-CMO) interface, we assume that the density
is uniform in each layer for calculating the density profile as a function
of distance $z$ from the insulator-oxide interface.
The Coulombic interaction energy per unit area due to leaked charges is
the same for both LMO and CMO regions and is given, in the continuum approximation,
to be
\begin{equation}
E_{\rm coul} = \frac{1}{8\pi \epsilon}\int_0^{L} dzD(z)^2 ,
\end{equation}
where $D(z)$ is the electric displacement and is given by
\begin{equation}
D(z) =
 \pm \left [-\epsilon {\rm E}_{\rm ext} + \int_0^z dy 4\pi e \rho(y) \right ] ,
\label{E_c}
\end{equation}
with $+$ ($-$) sign for LMO (CMO) side. Furthermore,
 $\rho(z)$ is the 
 density of minority charges (i.e., holes on LMO side and electrons on SMO side),
  $e$ the charge of a hole, 
$\epsilon$ the dielectric constant,
${\rm E}_{\rm ext}$  an external electric field along the z-direction,
and $L$  the thickness of the LMO (CMO) layers.

The ground state
energy per unit area [corresponding to the effective Hamiltonian
 of Eq. (\ref{H_pol})]
can be written, in the continuum approximation, as
 a functional of $\rho(z)$ for both LMO and CMO as follows:
\begin{equation}
E_{\rm pol} = -
\left [\gamma^1_{\rm ep} g^2\omega_0 +\frac{\gamma^2_{\rm ep}t^2}{g^2\omega_0} \right ]
\zeta\int_0^{L} dz
 \rho(z) [1- a^3  \rho(z)] ,
\label{E_pol}
\end{equation}
where $\zeta=6$ is the coordination number and $a$ is the lattice constant.
In arriving at the above equation 
we have approximated $n_{i+\delta}+n_{i-\delta} \approx 2n_i$.
Furthermore, for the ground state, we expect 
  $t_{ij}\sqrt{0.5[1+({\bf S}
 _i\cdot{\bf S}_j/{\bf S}^2)]} =t_{ij}$, i.e., the minority charges will completely
polarize the neighboring majority charges.

We will now minimize the total energy given below
\begin{equation}
E_{\rm Total} = 2 E_{\rm pol} +2 E_{\rm coul} ,
\label{E_Total}
\end{equation}
by setting the functional derivative $\delta E_{\rm Total}/\delta \rho(z)$ =0.
This leads to the following equation
\begin{eqnarray}
0 &=&-C_1 [1- 2 a^3  \rho(z)] 
+ 2 C_2\int_{z}^{L} dy \left [- {\tilde{\rm E}}_{\rm ext} + \int_0^y dx \rho(x) \right ] , 
\nonumber \\
&&
\label{E_opt}
\end{eqnarray}
where $C_1 \equiv \left [\gamma^1_{\rm ep} g^2\omega_0 +\frac{\gamma^2_{\rm ep} t^2}{g^2\omega_0} \right ] \zeta$;
 $C_2 \equiv 2\pi e^2/\epsilon $; and
$ {\tilde{\rm E}}_{\rm ext} \equiv \epsilon {\rm E}_{\rm ext}/(4 \pi e)$.
The above equation, upon taking double derivative with respect to $z$, yields
\begin{equation}
C_1 a^3 \frac{d^2 \rho(z)}{d z^2} -C_2 \rho(z)=0.
\label{rhoz_eq}
\end{equation}
The above second-order differential Eq. (\ref{rhoz_eq})
 and Eq. (\ref{E_opt})
 admit the solution
\begin{equation}
\rho(z)= \frac{1}{2 a^3} \frac{\cosh(\xi z)}{\cosh(\xi L)} +
\xi {\tilde {\rm E}}_{\rm ext}  \frac{\sinh[\xi (L-z)]}{\cosh(\xi L)} ,
\label{rhoz}
\end{equation}
where $\xi=\sqrt{C_2/(C_1 a^3)}$.
It is important to note that, for
{the manganese-oxide (MO)} layer
at the LMO-CMO interface (i.e., at $z=L$), the density is 0.5 electrons/site
and that it is 
{independent of the applied external electric field
and the system parameters. Now, since each ${\rm Mn}$ site
 in the interface layer belongs to a unit cell that is half LMO and half CMO, one expects the
 density per site to be 0.5. Additionally,  as the distance from the LMO-CMO interface increases,
 we observe from Eq. (\ref{rhoz}) 
as well as from Figs. \ref{fig:2layer_c1_0.24} and \ref{fig:2layer_c1_0.31} that for smaller
values of $C_1$, the density falls more rapidly while
the density change due to electric field rises faster. Furthermore, 
for realistic values of the parameters, the charge density 
rapidly changes as we move away
from the oxide-oxide interface (i.e., after only a few layers from the interface) and
attains values close
to the bulk value 
{[as illustrated in Figs. \ref{fig:2layer_c1_0.24}(a) and \ref{fig:2layer_c1_0.31}(a)]}. Lastly, as required for zero values
of the external field ${\rm E}_{\rm ext}$, we get $\rho(0) \rightarrow 0$ when  $L \rightarrow \infty$.
Thus, although we used the continuum approximation, our obtained density
profile is qualitatively realistic as it has the desired values at the extremes $z=L$
and $z=0$ with the density away from the LMO-CMO interface rapidly falling for not too large values of
$\epsilon$.}
\\

{
{The density profiles  for both LMO and CMO sides depend only on $\epsilon$,
 $C_1$ and ${\rm E}_{\rm ext}$.
{For our calculations displayed in Fig. \ref{fig:2layer_c1_0.24}, 
we used the following values for the parameters:
$a=4$ \AA; $\epsilon =20$; ${\rm E}_{\rm ext}= 300~{\rm kV/cm}$ and $400~{\rm kV/cm}$;
and  $C_1=0.24$. For Fig. \ref{fig:2layer_c1_0.31},
we employed ${\rm E}_{\rm ext}= 100~{\rm kV/cm}$ and $C_1=0.31$, with the values for $a$ and $\epsilon$
being the same as in Fig. \ref{fig:2layer_c1_0.24}.
The values of $C_1$ in Figs. \ref{fig:2layer_c1_0.24} and \ref{fig:2layer_c1_0.31} were
chosen based on $\omega_0 = 0.07 ~{\rm eV}$;
$g =2 $; and 
$ t = 0.1 ~{\rm eV}$.} 
}

\begin{figure}[b]
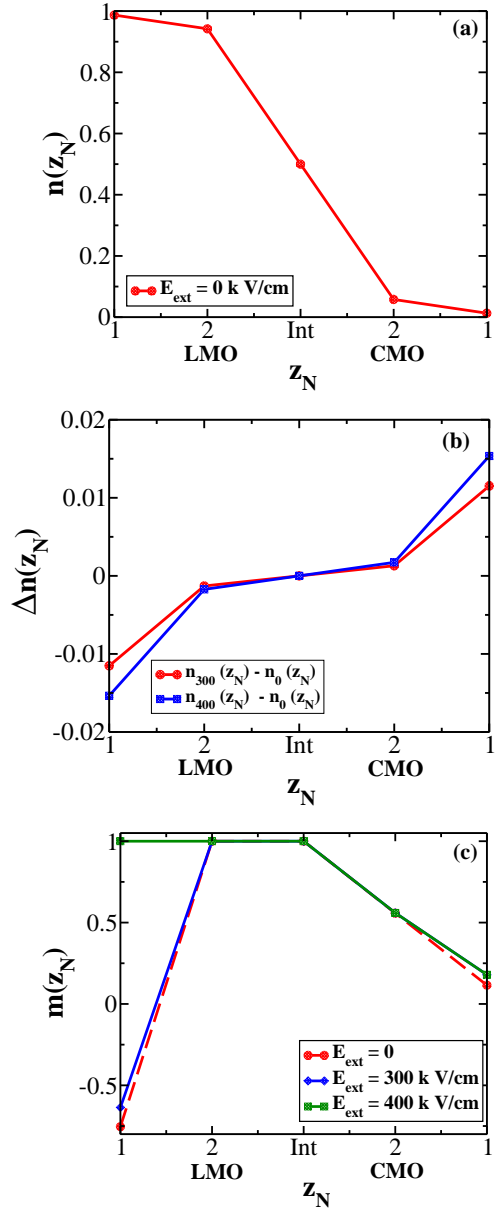

\includegraphics[width=6.0cm]{fig2a.eps}\\
\vspace{0.2cm}
\hspace{-0.5cm}
\includegraphics[width=6.45cm]{fig2b.eps}\\
\vspace{0.3cm}
\includegraphics[width=6.0cm]{fig2c.eps}
\caption{(Color online)
{
Electronic charge density $n(z_N)$ and 
{per-site magnetization  $m(z_N)$ (of $t_{\rm 2g}$ spins normalized to unity)}
 in various manganese-oxide layers 
of a ${\rm (Insulator)/(LaMnO_3)_2/(CaMnO_3)_2/(Insulator)}$ heterostructure
for  $a=4$ \AA, $\epsilon=20$, and $C_1= 0.24$.
Figures are for (a) $n ({\rm E}_{\rm ext} =0~{\rm kV/cm}) $; (b) $\Delta n = n ({\rm E}_{\rm ext}=300 /400~{\rm kV/cm} 
) - n ({\rm E}_{\rm ext}=0 ~{\rm kV/cm})$; and (c) $m(z_N)$ at 
${\rm E}_{\rm ext}=0 ~{\rm kV/cm},~300~{\rm kV/cm}$, and $400 ~{\rm kV/cm}$.
MO layer 1 on the LMO side undergoes spin reversal when  ${\rm E}_{\rm ext}=400 ~{\rm kV/cm} $ is applied.}}
\label{fig:2layer_c1_0.24}
\end{figure}

\begin{figure}[b]
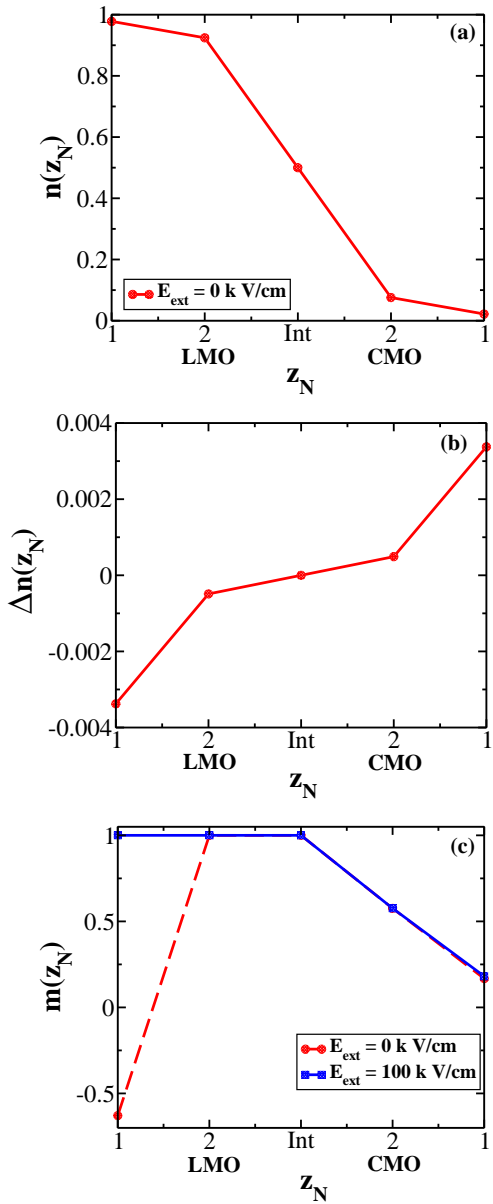

\includegraphics[width=6.0cm]{fig3a.eps}\\
\vspace{0.2cm}
\hspace{-0.5cm}
\includegraphics[width=6.45cm]{fig3b.eps}\\
\vspace{0.3cm}
\includegraphics[width=6.0cm]{fig3c.eps}
\caption{(Color online)
{
Electronic charge density $n(z_N)$ and 
{per-site magnetization  $m(z_N)$ (of normalized-to-unity $t_{\rm 2g}$ spins )}
 in various manganese-oxide layers 
of a ${\rm (Insulator)/(LaMnO_3)_2/(CaMnO_3)_2/(Insulator)}$ heterostructure
for  $a=4$ \AA, $\epsilon=20$, and $C_1= 0.31$.
Plots pertain to (a) $n ({\rm E}_{\rm ext} =0~{\rm kV/cm}) $; (b) $\Delta n = n ({\rm E}_{\rm ext}=100 ~{\rm kV/cm} 
) - n ({\rm E}_{\rm ext}=0 ~{\rm kV/cm})$; and (c) $m(z_N)$ at ${\rm E}_{\rm ext}=0 ~{\rm kV/cm}$ and
${\rm E}_{\rm ext}=100 ~{\rm kV/cm}$.
The LMO  side becomes completely ferromagnetic when  ${\rm E}_{\rm ext}=100 ~{\rm kV/cm} $ is applied.}}
\label{fig:2layer_c1_0.31}
\end{figure}

\subsection{Magnetization distribution}
\label{analytic_mag}
We will now obtain the magnetization for a 
 heterostructure by considering its lattice structure unlike
the case for the density profile where a continuum approximation was made. 
Thus we can take into account the possibility of 
{antiferromagnetic (AFM)} order
besides being able to consider 
{ferromagnetic (FM)} order.

{First, based on Ref. \onlinecite{cheong}, we note that the bulk ${\rm La_{1-x}Ca_xMnO_3}$  (below the
magnetic transition temperatures)
is A-AFM for $0 \le x \lesssim 0.1$  and a ferromagnet for $0.1 \lesssim x \lesssim 0.5$.
Hence, we model the LMO side
of the heterostructure 
as an A-AFM  when hole concentrations  are small; whereas at higher concentrations of holes which is
less than 0.5, the holes dictate the magnetic order by forming magnetic polarons
that polarize the A-AFM.
Next, we note that for $0.5 \lesssim x \lesssim 1.0$, the ${\rm La_{1-x} Ca_x MnO_3}$ bulk system is always an 
antiferromagnet. }
Thus, 
from a magnetism point-of-view, the
magnetic moment on the CMO side of the heterostructure is expected to be zero except in the vicinity 
of the interface where (due to proximity effect) it will be a ferromagnet and can be modeled using
a percolation picture. Given the above scenario, as can be expected, we find that the ferromagnetic
region on the LMO side can be drastically enhanced (at the expense of the A-AFM region)
by an electric field inducing holes on the LMO side. On the other hand, the electric field
has only a small effect on the percolating ferromagnetic cluster that is adjacent to the oxide-oxide
interface on the CMO side.

\subsubsection{CMO side}
We will first consider the CMO side and show that the magnetization
decays as we
move away from the LMO-CMO interface. We derive below the largest FM domain;
this domain percolates from the LMO-CMO interface.
 On account of nearest-neighbor
repulsion (as given in Eq. (\ref{H_pol})), the interface (which is half-filled)
has $e_{\rm g}$ electrons on alternate sites.
In fact, to minimize the interaction energy, on the CMO side we take the $e_{\rm g}$ electrons
to be in one sublattice only which  will be called $e_{\rm g}$-sublattice; the other unoccupied
sublattice  will be called  the $u$-sublattice.
On account of virtual hopping, an $e_{\rm g}$ electron polarizes all its neighboring
sites  that do not contain  any $e_{\rm g}$ electrons and forms a magnetic polaron.
Thus we observe that the half-filled interface  will be fully polarized and that
there will be an FM cluster that begins at the interface
and percolates to the layers away from the interface on the CMO side.
For instance, in the layer next to the oxide-oxide interface, all the  $e_{\rm g}$
electrons are in the same sublattice (the $e_{\rm g}$-sublattice)
and are next to the empty sublattice (i.e., sublattice  
unoccupied by $e_{\rm g}$ electrons) of the interface
 and hence are ferromagnetically aligned with the interface. Similarly, again in the 
 layer adjacent to LMO-CMO interface,
all the sites in the other sublattice (i.e., the $u$-sublattice) are empty and
have the same polarization
as the sites occupied by the $e_{\rm g}$ electrons at the interface.

We will now identify the equations governing the ferromagnetic cluster percolating from the interface.
Let $z_{\rm N}$ be the $z$-coordinate of the ${\rm N}^{\rm th}$ 2D MO layer
with ${\rm N}$ being the index measured from the 
Insulator-CMO  interface (as shown in Fig. \ref{fig:het_stut}). Furthermore, we define
$x^{e_{\rm g}}_{\rm N}$ ($x^u_{\rm N}$) as the concentration of polarized
sites that belong to the spanning cluster
and these sites are a subset of  the sites in the $e_{\rm g}$-sublattice ($u$-sublattice)
in the
${\rm N}^{\rm th}$ 2D MO layer.
Then, the factor
$(1-2x^{e_{\rm g}}_{\rm N})$ 
 [$(1-2x^{u}_{\rm N})$]
represents the probability of a site that belongs
to the $e_{\rm g}$-sublattice ($u$-sublattice)
in layer N
but is not part of
the spanning polarized cluster.
Now, the probability that a site, occupied (unoccupied)
by an $e_g$ electron, contributes to the FM cluster
is equal to the probability of finding the site occupied (unoccupied)
multiplied by 1-P where P is the probability 
that none of the adjacent sites 
that are in the 
$u$-sublattice
($e_{\rm g}$-sublattice)
belong to the percolating cluster. 
Therefore, we obtain the following set of coupled equations for the spanning cluster:
\begin{equation}
x^{e_g}_N =\rho(z_N)[1-(1-2x^{u}_{N-1})(1-2 x^u_N)^4(1-2x^{u}_{N+1})] ,
\label{magegN}
\end{equation}
\begin{equation}
x^{u}_N =0.5[1-(1-2x^{e_g}_{N-1})(1-2 x^{e_g}_N)^4(1-2x^{e_g}_{N+1})] .
\label{maguN}
\end{equation}
The boundary conditions involving layer 1 are
\begin{equation}
x^{e_g}_1 =\rho(z_1)[1-(1-2 x^u_1)^4(1-2x^{u}_{2})] ,
\label{mageg1}
\end{equation}
and 
\begin{equation}
x^{u}_1 =0.5[1-(1-2 x^{e_g}_1)^4(1-2x^{e_g}_{2})] ,
\label{magu1}
\end{equation}
while those for the LMO-CMO 
 interface are $x^{e_g}_{\rm Int}=x^{u}_{\rm Int} =0.5$.

\subsubsection{LMO side}
Next, we will show that the  LMO side with only a few layers,
for some realistic values of parameters,
can have a sizeable change in the magnetism when a large electric field is applied 
and thus can be exploited to obtain
a giant magneto-electric effect.
Similar to the 
{bulk situation in ${\rm La Mn O_3}$} \cite{khali,super_ex}, in our heterostructure
as well, we assume that  two spins
on any adjacent 
sites in each MO layer have a ferromagnetic coupling 
{ $J_{\rm xy} = 1.39 ~{\rm meV}$;}
  whereas, 
 any two neighboring spins
  on adjacent layers have  an
 antiferromagnetic coupling 
{ $J_{\rm z} = 1.0 ~{\rm meV}$.}
On the other hand, for  an electron and a hole 
on neighboring sites either in the same MO layer or in adjacent MO layers, (due to virtual hopping
of electron between the two sites) there is a strong
 ferromagnetic
coupling  
{$J_{\rm eh} = \gamma^2_{\rm ep} t^2/(g^2\omega_0) >> J_{\rm xy}$} \cite{tvr}.  
In our calculations, as long as the ratio of $J_{\rm xy}/J_{\rm z}$ is taken as fixed and $J_{\rm eh}$ is the
 significantly dominant coupling, we get the same magnetic picture.

In a LMO side with
a few MO layers, to demonstrate the possibility of large magnetization change upon the
application of a large external field,
we assume that the LMO-CMO interface  and all MO layers
up to layer M are completely polarized. Next, we assume that MO layers M and M-1
have low density of holes so that there is a possibility that
spins in MO layer M-1 are not aligned with the block of MO layers starting from
the LMO-CMO interface and up to 
layer M.
 We then analyze 
the polarization
of MO layer M-1  by  comparing the energies for the following two cases:
(i) layers M-1 and M are antiferromagnetically aligned
with the holes in layers M and M-1 inducing polarization only on  sites that are adjacent to the holes;
 and   
(ii) MO layer M-1 is completely polarized and aligned with Layer M. 

In a few-layered  heterostructure ${\rm (Insulator)/(LaMnO_3)_2/(CaMnO_3)_2/(Insulator)}$,
{for $C_1 = 0.24$ eV
 [as shown in Fig. \ref{fig:2layer_c1_0.24}(c)] and for $C_1 = 0.31$ [as in Fig. \ref{fig:2layer_c1_0.31}(c)],
 we  obtain a striking magneto-electric effect.}
For zero external field, when ${\rm M=2}$ is considered, case (i) (mentioned above)
 has lower energy, i.e., layer 1 
is antiferromagnetically coupled
to layer 2.
 On the other hand, when a strong electric field
  {($\sim 100~ {\rm kV}/{\rm cm}$ for $C_1 = 0.24$ and
$\sim 400~ {\rm kV}/{\rm cm}$ for $C_1 = 0.31$)}
is applied, MO layer 1 (due to increased density of holes)
becomes completely polarized and ferromagnetically aligned with the rest of the layers
(i.e., MO layer 2
and the oxide-oxide interface) on the LMO side. {
Thus, we get a giant magneto-electric effect!}
{We have considered $C_1$ values 
ranging from 0.24 to 0.31 and obtained 
magnetoelectric effect for various
threshold electric field values. Changing $C_1$ is physically 
equivalent to changing the effective nearest-neighbor electron-hole attraction.
Thus a smaller $C_1$ value of 0.24
indicates a lower effective nearest-neighbor electron-electron repulsion
which requires a larger
external electric field strength
of ${\rm 400~ kV/cm}$ to generate enough holes on the LMO side
and, consequently, flip the magnetization of layer 1. 
Obviously, for the larger value $C_1 = 0.31$, sufficient number of holes are already present
in the LMO system and layer 1 becomes
ferromagnetic when a smaller ${\rm E_{ext}~ (= 100 ~ kV/cm})$ is applied. At still larger
values of $C_1$, i.e., $C_1 \geq 0.32$, the LMO side is completely ferromagnetic even in the absence
of an external field.}

\section{Numerical Approach}
Here, in this section, we construct a detailed 2D model Hamiltonian and study it numerically
for the charge and magnetic profiles and the coupling between them.
\subsection{Model Hamiltonian}
{In a quasi 2D heterostructure (involving only a few 2D layers of
 both manganites), as mentioned in Sec. \ref{sec2},
we expect only a single narrow-width  polaronic band
to be relevant \cite{tvr}.
For our numerical treatment of a 2D lattice (with $l_1$ rows and $l_2$ columns),
we employ the following one-band Hamiltonian}:
\begin{eqnarray}
H = H_{\rm KE} + H_{\rm pol}^{\rm mf} + H_{\rm SE} + H_{\rm coul} +  H_{\rm V} .
\label{H_tot}
\end{eqnarray}
The kinetic energy term $H_{\rm KE}  $ is given by
\begin{eqnarray}
 H_{\rm KE} = -te^{-g^2}\sum\limits_{\langle i,j \rangle} 
\left [{\cos \left ({\theta_{ij}\over 2}\right ) {c^{\dagger}}_{i} c_{j}} + {\rm H.c.}\right ] ,
\label{H_ke}
\end{eqnarray}
where $t$ is the hopping amplitude that is attenuated by the electron-phonon coupling $g$ and
$c_{j}$ is the $e_{\rm g}$ electron destruction operator; furthermore,
$\cos (\theta_{ij}) $ is the modulation due to infinite Hund's coupling between the itinerant electrons
and the localized $t_{\rm 2g}$ spins  with $\theta_{ij}$ being the angle between 
two localized $S=3/2$  spins at sites $i$ and $j$ \cite{gennes,Izyumov}.
The second term $H_{\rm pol}^{\rm mf}$ in Eq. (\ref{H_tot}) is the mean-field version
of $H_{\rm pol}$ in Eq. (\ref{H_pol}) and is expressed as
\begin{eqnarray}
 H_{\rm pol}^{\rm mf} &=& \left [{{{\gamma}^1_{\rm ep} g^2 \omega_{0}}}   + 
{{\gamma}^2_{\rm ep} t^2 \cos^2 \left ({\theta_{ij}\over 2}\right ) \over {g^2 \omega_{0}}}\right ]\nonumber\\
&&~~ \times \sum\limits_{i,\delta}{\big [ n_i - 2n_i \langle n_{i+\delta}\rangle + \langle n_i \rangle 
\langle n_{i+\delta}\rangle \big ]} ,
\label{H_pol_mf}
\end{eqnarray}
where $\langle n_i \rangle \equiv \langle c^{\dagger}_i c_i \rangle$ refers to the mean number density at site $i$. 
A derivation of $H_{\rm KE} + H_{\rm pol}$ 
is given in Ref. \onlinecite{rpsy}; however, for simplicity, here we have ignored the effect
of next-nearest-neighbor hopping. The next term $H_{\rm SE}$ in Eq. ({\ref{H_tot}) pertains
to the superexchange \cite{anderson} term which generates A-AFM in $\rm LaMnO_3$ and G-AFM in $\rm CaMnO_3$; thus on the
$\rm LaMnO_3$ side, it is given by
\begin{eqnarray}
 H_{\rm SE}^{\rm lmo} = - J_{\rm x y} \sum\limits_{\langle i,j \rangle_{\rm x y}} \cos\left (\theta_{ij}\right )
 + J_{\rm z} \sum\limits_{\langle i,j \rangle_{\rm z}} \cos\left (\theta_{ij}\right ) ,
 \label{H_se1}
\end{eqnarray}
while on the $\rm CaMnO_3$ side, we  express it as
\begin{eqnarray}
 H_{\rm SE}^{\rm cmo} = 
 J_{\rm z} \sum\limits_{\langle i,j \rangle} \cos\left (\theta_{ij}\right ) .
 \label{H_se2}
\end{eqnarray}
In the above superexchange expressions, the magnitude of the $S=3/2$ spins is absorbed
in the superexchange coefficients $J_{\rm x y}$ and $J_{\rm z}$.
It should be clear that, away from the  LMO-CMO interface,
we can recover the spin arrangements of bulk $\rm LaMnO_3$ and  $\rm CaMnO_3$;
whereas, near the interface the minority carriers are expected to modify the 
spin textures.
In Eq. (\ref{H_tot}), the  Coulomb interaction is accounted for through the term  $H_{\rm coul}$ 
as follows:
\begin{eqnarray}
\!\!\!\!\!\! H_{\rm coul} = {{V_{\rm s}}}  \sum\limits_{i } n_i + \alpha t  \sum\limits_{i\neq j}\left [ n_i
\left ( {{\langle n_j\rangle - Z_j}\over {|\vec{r}_{i}-\vec{r}_j |}} \right ) 
- {{\langle n_i \rangle \langle n_j \rangle} \over {2 |\vec{r}_{i}-\vec{r}_j |}} \right ]  , \nonumber \\
\label{H_coul}
\end{eqnarray}
where the first  term on the right hand side (RHS) is actually
the on-site Coulomb interaction between an electron
and a positive ion that yields the binding energy of an electron
and is therefore  applicable only to the LMO side.
The remaining  term on the RHS of Eq. (\ref{H_coul}),  denotes long-range, mean-field Coulomb interactions 
between electrons as well as between
electrons and positive ions. 
Here,
$Z_j$ represents the positive charge density operator
with a value of either 1 or 0 and $|\vec{r}_i-\vec{r}_j|$ is the distance between lattice sites $i$ and $j$.
Furthermore, the dimensionless parameter  $\alpha =  {{e^2} \over {4\pi\epsilon a t}}$
determines the strength of the Coulomb interaction.
Lastly, in Eq. (\ref{H_tot}),  the term $H_{\rm V}$ represents the potential felt 
at various sites due to an
externally applied  potential  difference ($V_{\rm ext}$) between the two insulator edges (see Fig. \ref{fig:het_stut}):
\begin{eqnarray}
 H_{\rm V} = V_{\rm ext }\sum\limits_{\rm I=1}^{l_2} 
\sum\limits_{\rm K=1}^{l_1} \left [ 1- \left ({{\rm I-1}\over{l_2-1}} \right ) \right ] n_{ {{\rm I+(K-1)}l_2}} ,
\end{eqnarray}
where I represents the layer (or column) index 
with $l_2$ denoting the number of layers, i.e., the number of sites in the z-direction;
K represents the row index with
$l_1$ denoting the number of rows, i.e., the number of sites 
in a layer.

\subsection{Calculation procedure}
\label{calc_proc}
We consider a 2D lattice involving a few layers (columns) of LMO and CMO
and study magnetoelectric effect. 
The lattice does not have periodicity in
the direction normal to the interface of the heterostructure
(i.e., the z-direction). On the other hand, to mimic  
infinite extent in the 
direction parallel to the interface
of the heterostructure,
we assume periodic boundary condition in that direction.
We employ classical Monte Carlo  with Metropolis update algorithm
to obtain the  charge and magnetic profiles for our 2D lattice.
To tackle the difficult problem of several local minima that are close in energy, 
we take recourse to the simulated annealing technique.
To arrive at a reasonable charge profile at energy scales much
larger than the superexchange energy scale,
we treat the problem classically (i.e., fully electrostatically)
by considering only the Coulomb term ($H_{\rm coul}$), the external-potential term   ($H_{\rm V}$),
and the electron-phonon-interaction term  [$H_{\rm pol}^{\rm mf}(\theta_{ij} =0)$]
as these are the dominant energy
terms in the Hamiltonian.
The Coulomb term
is subjected to mean-field analysis 
(as mentioned before) and the system generated
potential $\alpha t \sum\limits_{i\neq j} \left \{
{{\langle n_j\rangle - Z_j}\over {|\vec{r}_{i}-\vec{r}_j |}} \right \}$
\cite{millis}
in Eq. (\ref{H_coul})
is solved self-consistently.
This is equivalent to solving the Poisson equation\cite{dagotto2}.

Next, to arrive at the final charge and magnetic configurations,
we treat the system quantum mechanically by
starting with an initial configuration comprising of
the charge configuration generated classically (by the above procedure)
and an initial random spin configuration.
{We now consider the full Hamiltonian,
where hopping term ($H_{\rm KE}$) and spin interaction energy 
act as  perturbation
to the classical dominant energy terms 
[$H_{\rm coul}$, $  H_{\rm V}$, and $H_{\rm pol}^{\rm mf}(\theta_{ij} =0)$],
thereby allowing for a small change in the number density profile and 
determine the concomitant magnetic  profile.}
For the classical $t_{\rm 2g}$ spins
$\vec{S_i} = (\sin\theta_i \cos\phi_i, \sin\theta_i \sin\phi_i, \cos\theta_i)$ that are
normalized to unity,
the $\cos(\theta)$ and $\phi$ values are binned 
in the intervals $(-1,1)$ and $(0,2\pi)$ respectively with equally spaced 40 values of $\cos(\theta)$ 
and 80 values of $\phi$, hence yielding a total of 3200 different possibilities.

In our calculations, we  employ the parameter values 
$t=0.1~{\rm eV}$,  $g=2~\&~2.2$, and $\omega_0 = 0.07$ leading 
to a  small parameter value ${{t}\over{\sqrt{2} g\omega_0}} < 1$.
For our manganite heterostructure, lattice constant $a=4$ \AA, dielectric constant $\epsilon=20$,
magnetic couplings 
{ $J_{\rm z} = 1.00 ~{\rm meV}$ and $J_{\rm xy}/J_{\rm z} = 1.39$;}
we take the pre-factors [in Eq. (\ref{H_pol_mf})] $\gamma^1_{ep} = 0.3$ and $\gamma^2_{ep} = 0.25$. 
The coefficient  in Eq. (\ref{H_coul}),  representing the
relative work function of the LMO side with respect to the CMO side, is taken to be $V_{\rm s}= 3\alpha$; 
hence the confining radius for the $e_{\rm g}$ electron is $1.33$~\AA~  which is less 
than half the lattice constant.
{External potential differences $V_{\rm ext}$, corresponding to 
external electric fields ${\rm E_{ext}} = 300 ~{\rm kV/cm}$ and ${\rm E_{ext}} = 400~ {\rm kV/cm}$ 
(which are less than the breakdown field in LCMO \cite{ramesh}), 
are applied to study changes in the magnetization profiles}.

The simulation (involving the charge and spin degrees of freedom)
is carried out
using exact diagonalization
of the total Hamiltonian in Eq. (\ref{H_tot}).
The spins 
are annealed over 61 values of the dimensionless temperature [$k_B T/(te^{-g^2})$],  
in steps of 0.05,
starting from 3 
and ending at 0.05
 with 15000 system
sweeps carried out at each temperature.
Since hopping energy $t e^{-g^2} > J_{\rm xy}$,
 the inclusion 
of the spin degrees of freedom certainly commences
at temperatures
$k_B T > J_{\rm xy}$.
Furthermore, the endpoint $k_B T=0.05te^{-g^2}$ is  sufficiently
small to correspond to the ground state of the system.
Each sweep requires visiting 
all the lattice sites  sequentially and updating
the spin configuration at each lattice site
by the standard Metropolis Monte Carlo algorithm.
We are also allowing the charge degrees of freedom
to relax by treating the problem self-consistently.
So, at the  beginning of each sweep,
the Poisson equation is solved additionally 
 to make sure that the number densities have
converged and this is achieved with an accuracy of 0.001.
Finally, averages of the various measurables
in the system are taken over the last 5000 sweeps in the system.

\subsection{Results and discussion}
\label{res_disc}
For numerical simulation,
we consider two  lattice sizes, namely, $12\times6$ and $12\times8$
with number of rows $l_1 =12$ and number of layers (columns) $l_2 =6$ or $8$.
Here, all the ${\rm Mn}$ sites in each layer
belong solely to either LMO or CMO.
This is in contrast to the continuum approximation employed in Sec. \ref{analytic_charge}
to obtain the charge profile analytically.
In Sec. \ref{analytic_charge}, by exploiting the symmetry of the interactions of the minority
carriers on both sides of the LMO-CMO interface, we derived the charge profile with 
charge density always $\langle n \rangle=0.5$   at the interface; this corresponds to a system
comprising of odd number of ${\rm MnO_2}$ layers with the interface ${\rm MnO_2}$ layer being shared equally by 
 the LMO and CMO sides, i.e., 
 each ${\rm Mn}$ site (in the interface ${\rm MnO_2}$ layer)
belongs to a unit cell that is half LMO and half CMO.

We consider various situations in our lattices. First, we analyze the case of excluding
electron-phonon interaction; consequently, the  Hamiltonian of interest
is that given by Eq. (\ref{H_tot}), but without the $H_{\rm pol}^{\rm mf}$ term. 
Next, we study the charge and magnetic profiles predicted by the total Hamiltonian of  Eq. (\ref{H_tot})
for the symmetric situation (of equal number of LMO and CMO layers) and
for different 
{sizeable} values of the electron-phonon coupling, i.e., for $g=2~\&~2.2$.
Lastly, we examine the impact on the magnetoelectric effect due to the asymmetry in number of LMO and CMO layers.

\subsubsection{No electron-phonon interaction and ${\rm E_{ext}} = 400 ~{\rm kV/cm}$}

\begin{figure}[b]
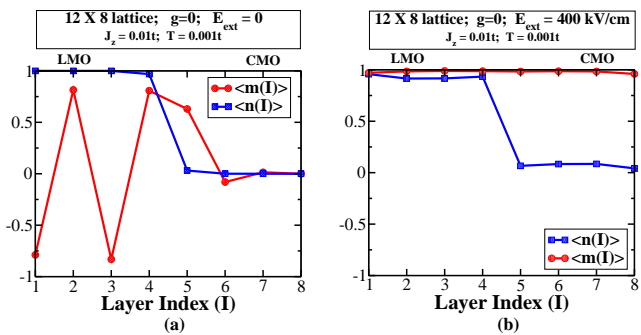

 \includegraphics[width=4.0cm]{fig4a.eps} ~~ 
 \includegraphics[width=4.cm]{fig4b.eps} \\
\caption{(Color online) 
{Layer-averaged charge density $\langle n(I) \rangle $ and
layer-averaged per-site magnetization $\langle m(I) \rangle $
(of $t_{\rm 2g}$ spins normalized to unity)
for a symmetric $12\times8$ LMO-CMO lattice when electron-phonon interaction is zero; $J_z = 0.01t$ and
$J_{\rm xy}/J_{\rm z} = 1.39$; $T = 0.001t$;
and when (a) external electric field $E_{\rm ext} =0$ and 
(b) $E_{\rm ext} = 400~ {\rm kV/cm}$.}}
\label{fig:ep01}
\end{figure}

Here, without the electron-phonon interaction, the hopping amplitude $t$ is not attenuated
by the factor $e^{-g^2}$ and the ground state is a superposition of various states
in the occupation number representation. The electrons are not localized and we do not
need to employ simulated annealing; the calculations were performed at
a single temperature $k_B T = 0.001t$ on a symmetric heterostructure defined
on a lattice with equal number of layers on the LMO side  and the CMO side.
Furthermore,
 the charge density profile is essentially dictated by the
 Coulombic term in Eq. (\ref{H_tot}); the kinetic term and the superexchange term
have a negligible effect. Thus,
 we have density close to  1 on the LMO side and an almost zero density on the CMO side.
  Now, when a large electric 
 field (400 ${\rm kV/cm}$) is applied, a small amount of charge 
 gets pushed across the interface. Then, since the kinetic term is much larger
 than the superexchange term, double exchange 
  tries to ferromagnetically align the spins
 and, as  shown in 
  {Fig. \ref{fig:ep01}}, 
 we get a large change in the total  magnetization 
 of $t_{\rm 2g}$ spins 
 of the system, i.e., 
  {
  0.91/site for the $12\times8$ lattice 
 when $t_{\rm 2g}$ spins are normalized to unity. 
 
 \begin{figure}[t]
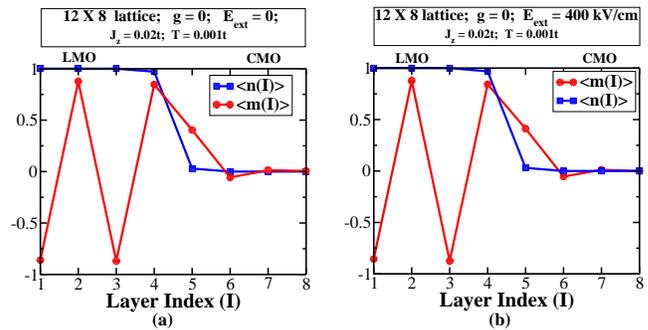

 \includegraphics[width=4.0cm]{fig5a.eps} ~~ 
 \includegraphics[width=4.cm]{fig5b.eps} \\
\caption{(Color online) 
{Layer-averaged charge density $\langle n(I) \rangle $ and
layer-averaged per-site magnetization $\langle m(I) \rangle $
(of $t_{\rm 2g}$ spins normalized to unity)
for a symmetric $12\times8$ LMO-CMO lattice when electron-phonon interaction $g=0$; superexchange $J_z = 0.02t$
and coupling ratio $J_{\rm xy}/J_{\rm z} = 1.39$; 
$T = 0.001t$;
and when (a)  $E_{\rm ext} =0$ and 
(b) $E_{\rm ext} = 400~ {\rm kV/cm}$.}}
\label{fig:ep02}
\end{figure}
 
  {Keeping the temperature fixed at 0.001t, if we now double the
 value of $\rm J_z$ to 0.02t while retaining the magnetic-coupling 
 ratio $J_{\rm xy}/J_{\rm z} = 1.39$, it is found that that the magnetoelectric effect
  disappears completely. 
 Owing to the larger superexchage interaction, there is only a small change in
 the density on the LMO and CMO sides. Consequently,  superexchange dominates
 over  double exchange, thereby
  making the system totally
 antiferromagnetic (i.e., similar to the bulk, the LMO side is A-AFM and the CMO side is G-AFM).
 Furthermore, even after the application
 of a large external electric field (i.e., ${\rm E_{ext}} = 400 ~{\rm kV/cm}$),
 there is practically no change in the magnetization 
  as demonstrated in {Fig. \ref{fig:ep02}}}. 
  
  {
 Next, 
  on retaining the superexchange interaction values  of $J_z = 0.01t$ and
$J_{\rm xy}/J_{\rm z} = 1.39$, when
  the temperature is increased from 0.001t to 0.01t, the disordering effect of the
  temperature dominates over superexchange making the magnetic profile 
  lose its oscillatory nature on the LMO side [as can be seen by comparing
  Fig. \ref{fig:ep03}(a) with Fig. \ref{fig:ep01}(a)]. 
On the application of a sizeable
 external electric field (i.e., ${\rm E_{ext}} = 400 ~{\rm kV/cm}$), 
 minority carrier density increases on both LMO and CMO sides.
 However, the disordering effect of the enhanced temperature 
 diminishes the double exchange effect, thereby producing
 only a modest increase in the magnetization on both the LMO and CMO sides
 [see Fig. \ref{fig:ep03}(b) and Fig. \ref{fig:ep01}(b)].}

 \begin{figure}[t]
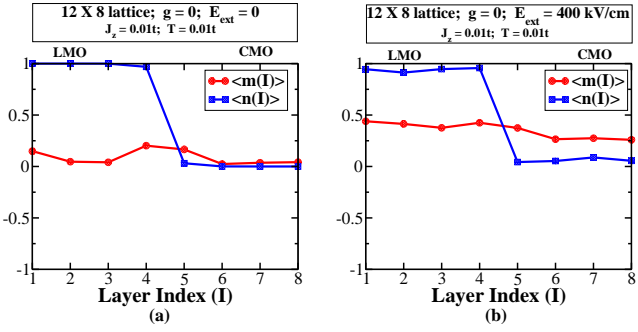

 \includegraphics[width=4.0cm]{fig6a.eps} ~~
 \includegraphics[width=4.cm]{fig6b.eps} \\
\caption{(Color online) 
{
Layer-averaged charge density $\langle n(I) \rangle $ and
layer-averaged per-site magnetization $\langle m(I) \rangle $
(of normalized-to-unity $t_{\rm 2g}$ spins )
for a symmetric $12\times8$ LMO-CMO lattice when $g=0$;  $J_z = 0.01t$ and
$J_{\rm xy}/J_{\rm z} = 1.39$;
 enhanced temperature $T = 0.01t$;
and when (a)  $E_{\rm ext} =0$ and 
(b) $E_{\rm ext} = 400~ {\rm kV/cm}$.}}
\label{fig:ep03}
\end{figure}
 
 In manganites the electron-phonon interaction is quite strong
and leads to sizeable cooperative oxygen octahedra
 distortions. Hence, to get a more 
 realistic picture, we switch on this interaction and study its effect on the system.
 Then, the hopping amplitude $t$ is attenuated by the factor $e^{-g^2}$
 and the electrons are essentially localized.
 Consequently, the states  are more or less classical in nature with number density
 at each site close to 1 (i.e., $> 0.99$ from our calculations) or close to 0 (i.e., $< 0.01$
 from our numerics);
 the  state of the system can be represented by
 a single state in the occupation number representation.
 As discussed in 
 Sec. \ref{calc_proc}, we employ simulated annealing; we arrive at the
 charge and magnetic profiles reported in the subsequent Secs. \ref{12x8_g2_e300}--\ref{asym_6_2_12x8_g2_e300}}.

\subsubsection{Symmetric $12\times8$ lattice with $g=2.0$ and ${\rm E_{ext}} = 300 ~{\rm kV/cm}$}
\label{12x8_g2_e300}
\begin{figure}[!htb]
\includegraphics[width=5.0cm]{fig7a.eps} 
\includegraphics[width=3.4cm]{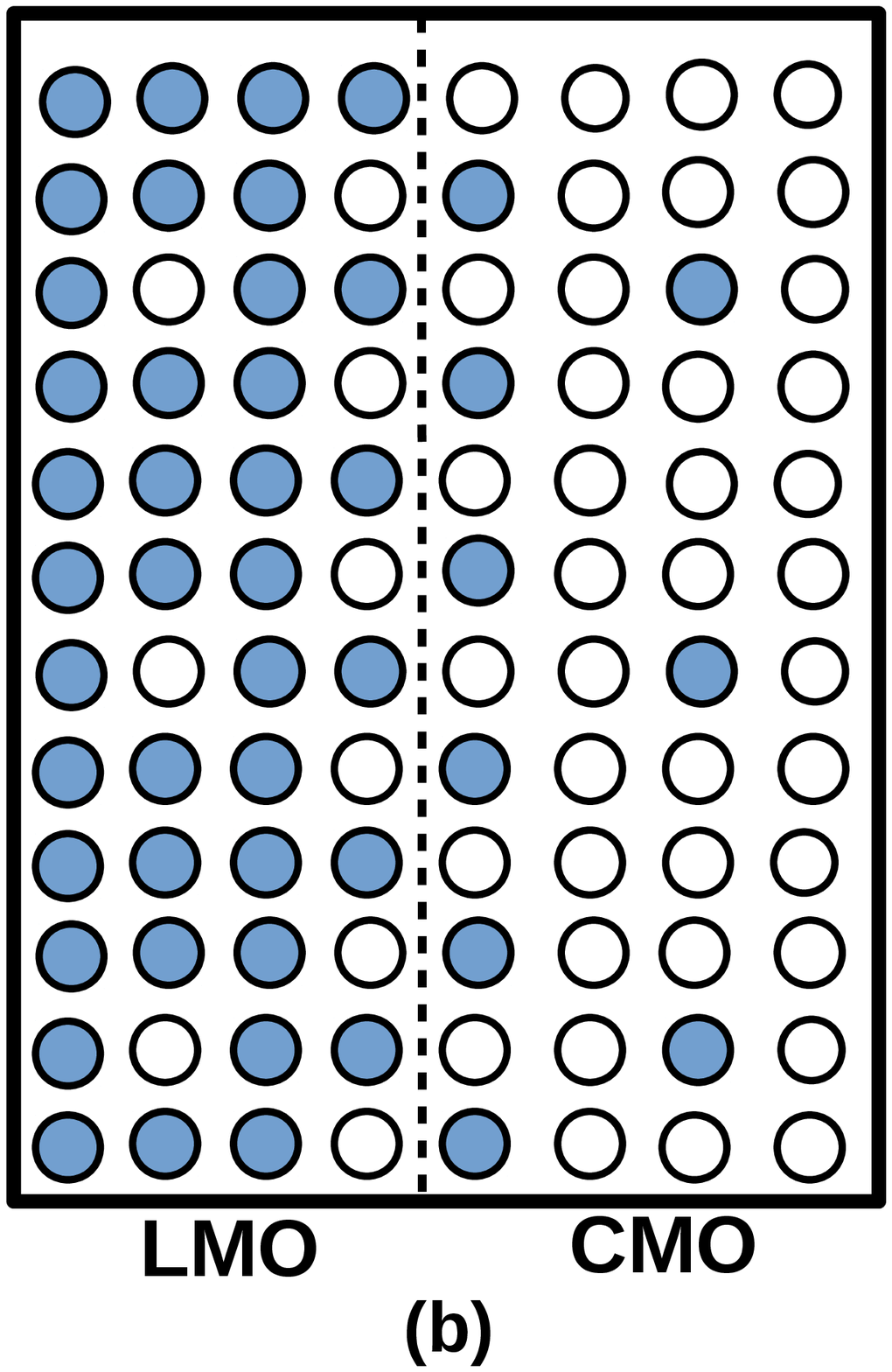} \\
\vspace{0.5cm}
\includegraphics[width=5.0cm]{fig7c.eps} 
\includegraphics[width=3.4cm]{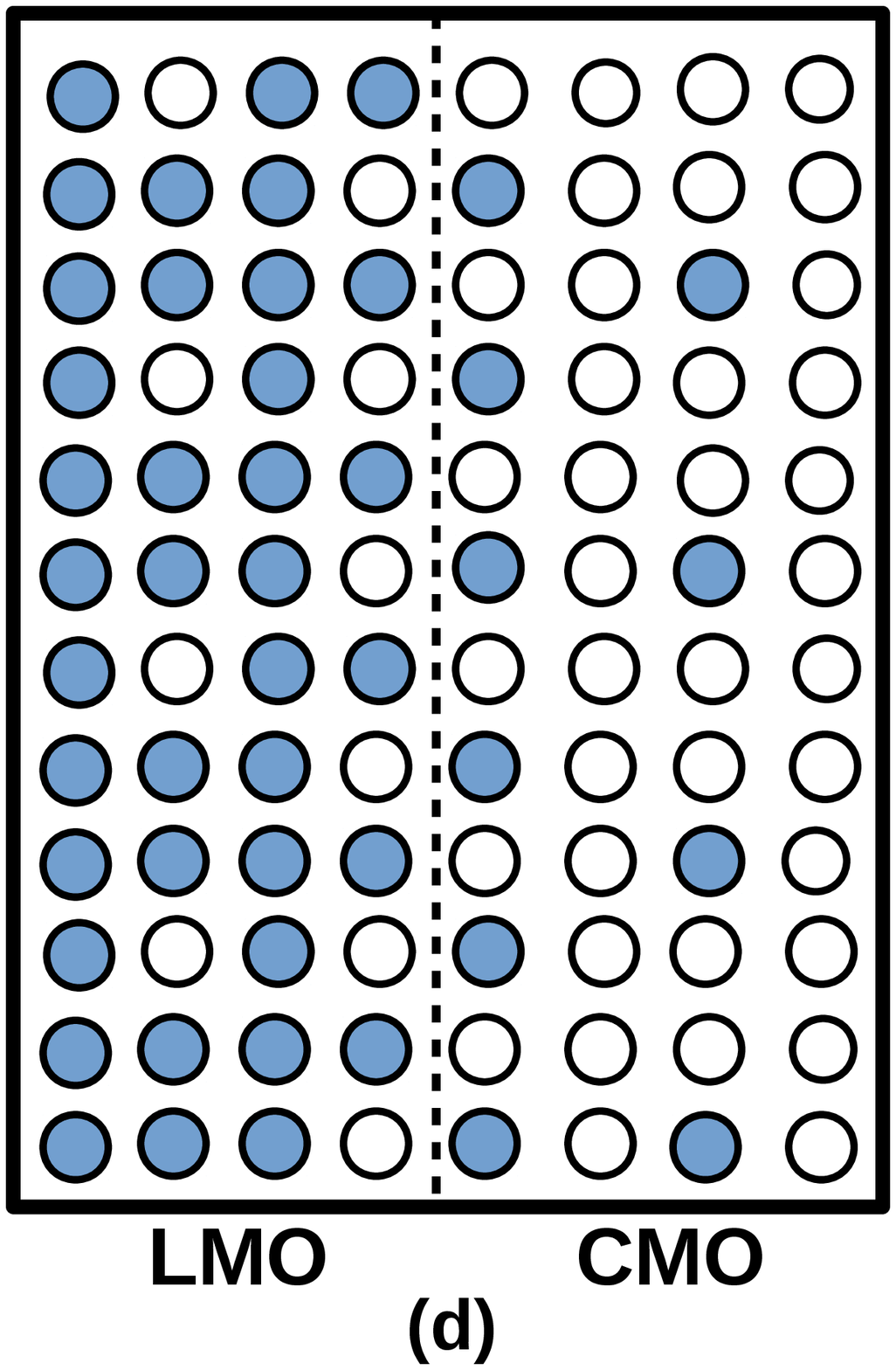} \\
\caption{(Color online) In a $12\times8$ symmetric lattice, when electron-phonon interaction strength $g=2.0$,
(a) when external electric field $E_{\rm ext} = 0$,
layer-averaged profiles of charge density $\langle n(I) \rangle $ and
magnetization $\langle m(I) \rangle $ (of the $t_{\rm 2g}$ spins normalized to unity); 
(b) when external electric field $E_{\rm ext} = 0$, schematic occupation-number representation of 
ground state charge configuration in the lattice;
(c) when a large external electric field $E_{\rm ext} = 300~ {\rm kV/cm}$ is applied,
modified layer-averaged charge density $\langle n(I) \rangle $ and layer-averaged
magnetization $\langle m(I) \rangle $ (of the $t_{\rm 2g}$ spins normalized to unity) for various layers 
in the lattice; and (d)  when $E_{\rm ext} = 300~ {\rm kV/cm}$,
reorganized ground state charge configuration.
}
\label{ep50_0}
  \end{figure}

We now consider a symmetric $12\times8$ lattice with 4 layers of LMO and another
4 layers of CMO as shown in Fig. \ref{ep50_0}.
We find charge modulation in the z-direction on both the sides, with neutral layers (free of minority carriers)
sandwiched between charged layers (with minority carriers). The layers at the interface 
have the largest number of minority carriers with electrons and holes on alternate sites
since contributions from both $H_{\rm pol}$  and $H_{\rm coul}$ [given by Eqs. (\ref{H_pol}) and (\ref{H_coul})] 
are minimized
for this arrangement.
Layers 1 and 8, being the farthest from the LMO-CMO interface, are devoid of any minority carriers
and retain the expected bulk charge distribution of LMO and CMO.
\begin{figure}[!htb]
 \includegraphics[scale=0.45]{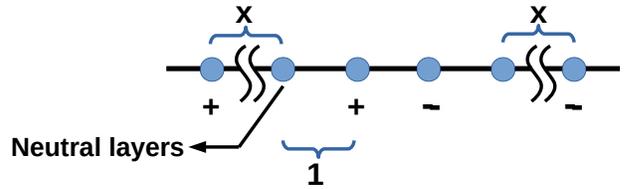}
 \caption{(Color online) Charge modulation due to Coulomb interaction $H_{\rm coul}$
 in a one-dimensional symmetric LMO-CMO lattice. The number of neutral layers/sites is $x$.}
  \label{charge_mod}
\end{figure}
We will now explain the charge modulation as follows.
We compute the energy $E_{\rm coul}$ electrostatically for the one-dimensional chain in Fig. \ref{charge_mod}
using $H_{\rm coul}$ in Eq.(\ref{H_coul}), 
with lattice constant taken as unity and the number of the neutral layers/sites as $x$.
Then 
\begin{align}
 E_{\rm coul} =& {1\over{x+1}}-{1\over{x+2}}-{1\over{2x+3}}-1-{1\over{x+2}}+{1\over{x+1}}\nonumber\\
 =& {2\over{x+1}}-{2\over{x+2}}-{1\over{2x+3}}-1 .
  \label{E_coul}
\end{align}
If we plot $ E_{\rm coul}$ as a function of $x$,
we find that it drops rapidly till $x=1$ and attains its minimum
value gradually somewhere between $x=6$ and $x=7$. 
Similarly, we expect neutral layers to be present in 2D also
and conclude that the charge ordering
sets in due to electrostatic Coulomb energy minimization.

\begin{figure}[t]
\includegraphics[width=6cm]{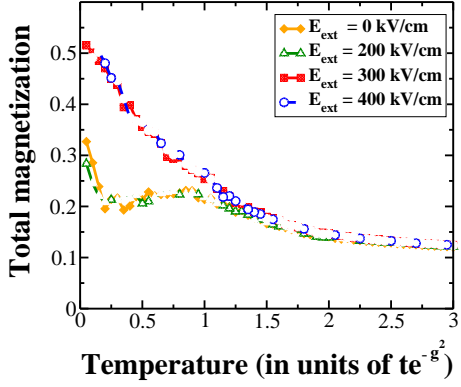}
\caption{(Color online) 
{Total magnetization as a function of temperature 
for various $E_{\rm ext}$ in a $12\times8$ lattice  
when electron-phonon coupling  $g=2.0$. Figure shows that an enhancement in magnetization occurs for
electric fields  $E_{\rm ext} \gtrsim 300$ kV/cm; whereas below this threshold value, total magnetization
does not change from its value at $E_{\rm ext} = 0$ kV/cm. The magnetoelectric effect is 
reasonably large at temperatures below 0.5t$e^{-g^2}$ ($\sim 10$ K).}
}
\label{mag_elec_T}
\end{figure}

In Fig. \ref{ep50_0},
the interface is fairly  polarized since 
the arrangement of electrons and holes on alternate sites produces a  strong ferromagnetic
coupling between  the spins on these sites as $J_{\rm eh} >> J_{\rm xy} > J_{\rm z}$.
Furthermore, again due to ferromagnetic couplings $J_{\rm eh}$ and $J_{\rm xy}$
on the LMO side, 
the interfacial layer polarizes the neutral layer 3 adjacent to it. 
Layer 1 is polarized in the direction of layer 3 due to antiferromagnetic coupling $J_{\rm z}$.
On the  CMO side, layer 6 is antiferromagnetic based on the charge configuration;
layer 8, as expected, is also fully antiferromagnetic.
As regards the case of zero electric field shown in  Figs. \ref{ep50_0}(a) and \ref{ep50_0}(b),
since layer 3 is antiferromagnetically connected to layer 2,
 layer 2 shows a small negative magnetization with the magnitude
diminished due to the presence of a few (i.e., 3)  holes in this layer.
On the application of a large electric field ${\rm E_{ext}} = 300 ~{\rm kV/cm}$,
as displayed in Figs. \ref{ep50_0}(c) and \ref{ep50_0}(d),
number of minority carriers increases in both layer 2 and layer 7.
Consequently, the magnetization increases in these layers. 
On the application of the external electric field, there is an overall
increase in the magnetization 
(of normalized-to-unity $t_{\rm 2g}$ spins) by 0.17/site leading to a 
giant magnetoelectric effect 
{(as can be seen in Fig. \ref{mag_elec_T})}

{A study of the total magnetization with temperature for various 
external electric fields (i.e., $\rm E_{ext}$  increased in steps of 100 kV/cm from 0), 
as depicted in Fig. \ref{mag_elec_T}, reveals that we need a
threshold field  $\simeq \rm 300~ kV/cm$
to get a fairly large increase in the total magnetization.
Only above the threshold value,
the density of minority charges increases and the resulting charge configuration
is modified; 
correspondingly, the  spin configuration gets altered too. Above $\rm 300 ~ kV/cm$,
the charge configuration gets frozen for consecutive higher electric fields up to $\rm 600 ~kV/cm$ 
and no change in the magnetic profile
can be expected. Although it may seem that much higher electric fields will further change the magnetization,
they will actually produce a breakdown.}

\subsubsection{
{Symmetric $12\times8$ lattice with $g=2.2$ and ${\rm E_{ext} = 0}$}}
\begin{figure}
[!htb]
 \includegraphics[width=5.0cm]{fig10a.eps} 
 \includegraphics[width=3.4cm]{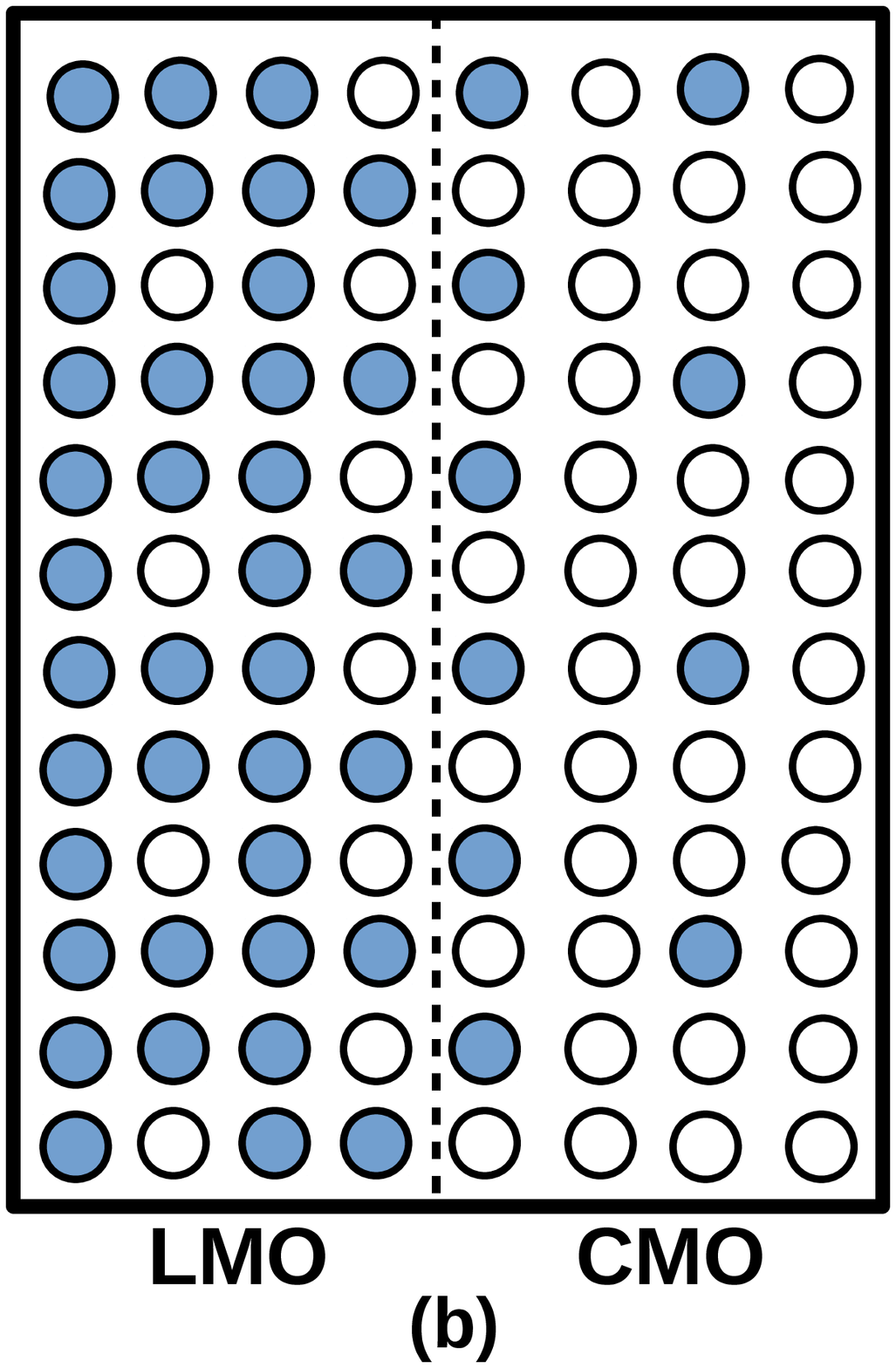} \\
\caption{(Color online)
Result of enhanced electron-phonon coupling $g=2.2$ and zero electric field, in  a symmetric $12\times8$ lattice,
on (a) layer-averaged charge density
$\langle n(I) \rangle $ 
and layer-averaged magnetization $\langle m(I) \rangle $ (of the $t_{\rm 2g}$ spins normalized to unity); and
(b) ground state charge configuration.}
\label{ep60_0_12x8}
  \end{figure} 
  
  Here we would like to point out that charge and magnetic
profiles, when  electron-phonon coupling $g$ is strong and 
 external electric field is zero, are similar to the profiles
 when coupling $g$ is weak
and external electric field is strong. In fact, as can be seen from Fig. \ref{ep60_0_12x8}
and Figs. \ref{ep50_0}(c) and \ref{ep50_0}(d),
for $g=2.2$ and ${\rm E_{ext} = 0}$ we
get the same charge profile as when $g=2.0$ and ${\rm E_{ext}} = 300 ~{\rm kV/cm}$;
the corresponding magnetic profiles in the two cases differ slightly because the ferromagnetic
coupling values $J_{\rm eh} = \gamma^2_{\rm ep} t^2/(g^2\omega_0)$ are slightly different. 

\subsubsection{Symmetric $12\times6$ lattice with $g=2.2$ and ${\rm E_{ext}} = 400 ~{\rm kV/cm}$}
\label{12x6_g2.2_e400}

\begin{figure}[t]
 \includegraphics[width=5.0cm]{fig11a.eps} 
 \includegraphics[width=2.7cm]{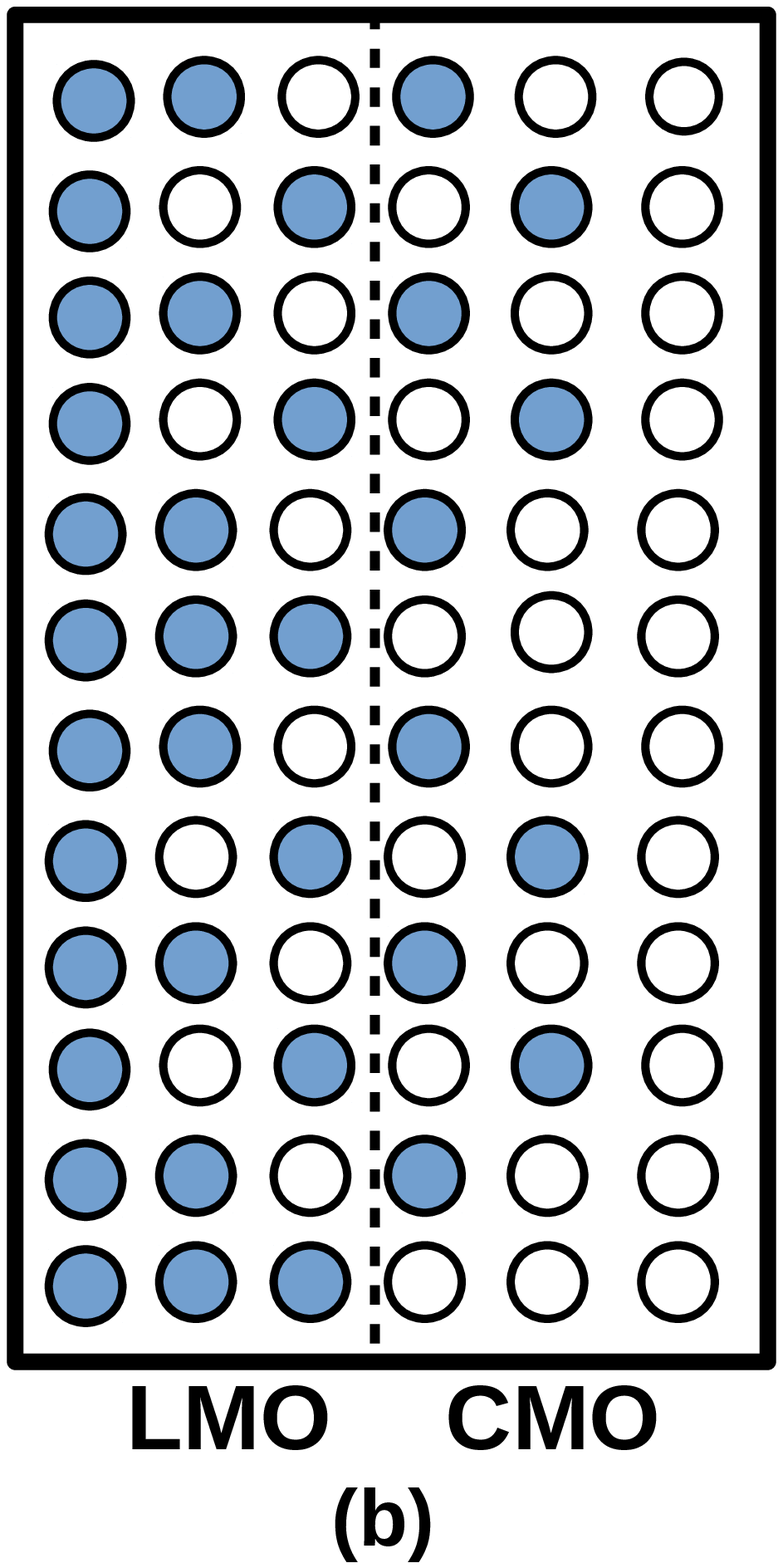} \\
 \vspace{0.5cm}
 \includegraphics[width=5.0cm]{fig11c.eps} 
 \includegraphics[width=2.7cm]{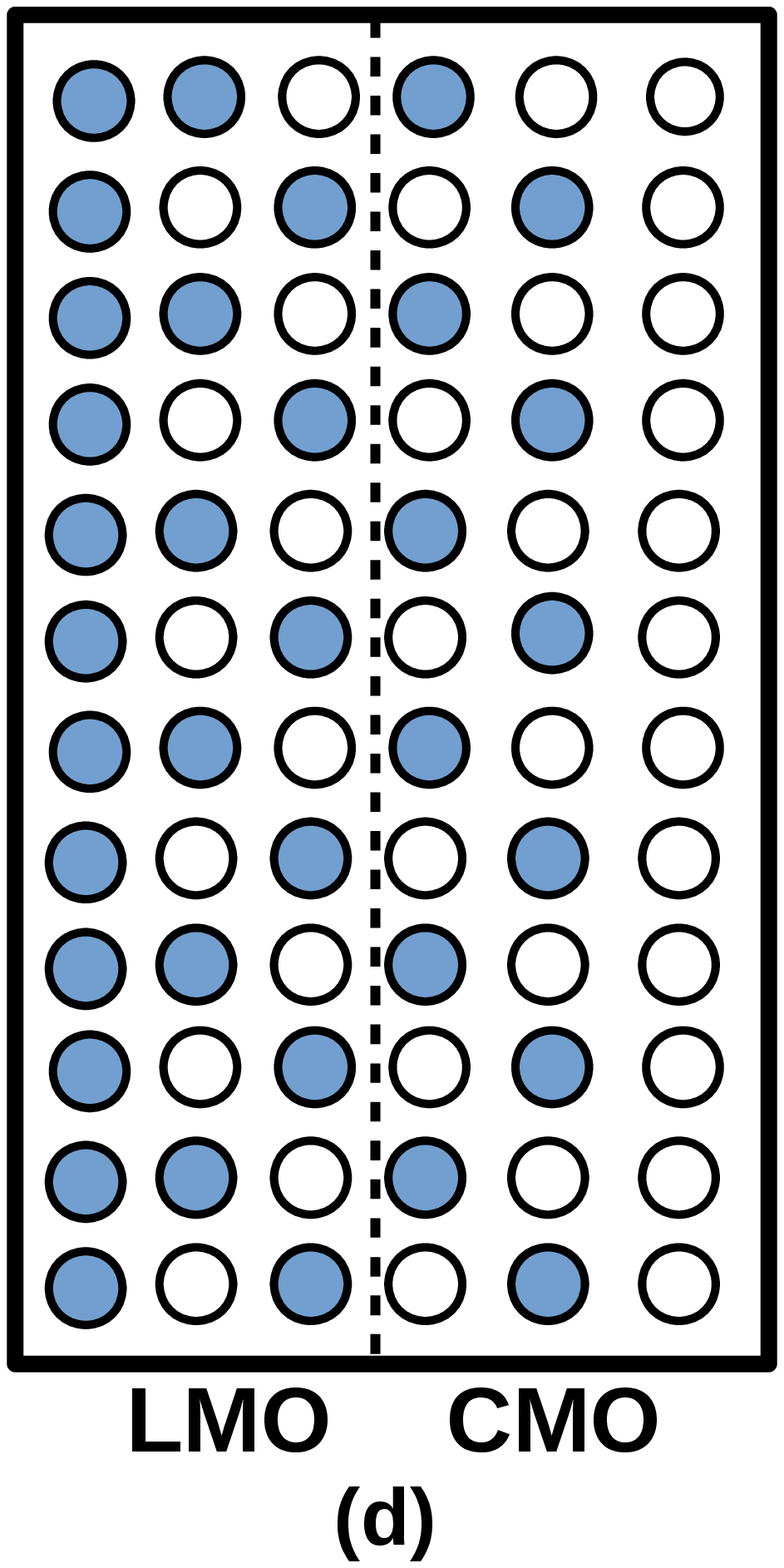} \\
\caption{(Color online)
In a symmetric $12\times6$ lattice, for enhanced coupling $g=2.2$, 
(a) at zero electric field, layer-averaged charge density $\langle n(I) \rangle $ and 
layer-averaged magnetization $\langle m(I) \rangle $ of $t_{\rm 2g}$ spins normalized to unity; 
(b) at ${\rm E_{ext}} = 0$, ground state configuration;
(c) at strong external electric field ${\rm E_{ext}} = 400 ~{\rm kV/cm}$,
layer-averaged charge density $\langle n(I) \rangle $ and 
layer-averaged magnetization $\langle m(I) \rangle $ of $t_{\rm 2g}$ spins normalized to unity; and
(d) at ${\rm E_{ext}} = 400 ~{\rm kV/cm}$, charge configuration in the ground state.
}
\label{ep60_0}
  \end{figure}

Since the electron-phonon interaction is stronger here (i.e., $g=2.2$) compared to the situation
in Sec. \ref{12x8_g2_e300},
even without the application of an external electric field, the concentration of minority
carriers is higher  (on both LMO side and CMO side) 
as can be seen from 
Figs. \ref{ep60_0} and \ref{ep50_0}.
Again, as depicted in Figs. \ref{ep60_0}(a) and \ref{ep60_0}(b),
the interfacial layers (i.e., layers 3 and 4) have electrons and holes
on alternate sites resulting in full ferromagnetism.
Here, we have fewer layers compared to the case
in Sec. \ref{12x8_g2_e300}; there is no charge modulation in the z-direction
because shifting the holes from layer 2 to layer 1 increases the number of  nearest-neighbor
repulsions due to electron-phonon interactions.
Layers 2  and 5 also have a sizeable density  of minority carriers (i.e., 1/3).
Consequently, layer 2 is polarized due to its proximity to layer 3 and
the ferromagnetic couplings $J_{\rm eh}$ and $J_{\rm xy}$;
contrastingly, the combination of $J_{\rm eh}$ and the  antiferromagnetic coupling
$J_{\rm z}$ lead to a smaller polarization in layer 5.
Layer 1, due to its proximity to layer 2 (with sizeable concentration of holes)
is also partially polarized.
On the application of a large electric field (${\rm E_{ext}} = 400 ~{\rm kV/cm}$), 
as portrayed in Figs. \ref{ep60_0}(c) and \ref{ep60_0}(d),
the concentration of  holes further increases in layers 2 and 5 leading
to fully ferromagnetically aligning these layers with layers 3 and 4.
Furthermore, layer 1 also gets more polarized.
On the whole, total magnetization (of normalized-to-unity $t_{\rm 2g}$ spins)
increases by $\sim 0.17$/site thus producing a giant magnetoelectric effect.

Lastly, it should be noted that (as expected) the 
magnetic profiles obtained
here in Fig. \ref{ep60_0}
are quite
similar to those in 
{Figs. \ref{fig:2layer_c1_0.24} and \ref{fig:2layer_c1_0.31};
on the other hand, the agreement between the charge profiles is not as good. }
{On comparing the analytic treatment with the numerical approach,
we note that the former 
makes a continuum approximation
to obtain the charge profile.
If the number of layers is large compared
to the lattice constant, the continuum approximation is valid 
and the prediction of the analytic approach will agree with the more accurate
numerical one.  
On the other hand, for a small system such as
a $12 \times 8$ system, 
 charge modulation is generated in the numerical approach unlike  the analytic case.
}

 \subsubsection{Asymmetric $12\times8$ system of 2  LMO layers and 6 CMO layers with $g=2.0$ and ${\rm E_{ext}} = 300 ~{\rm kV/cm}$ }
 \begin{figure}
 [t]
  \includegraphics[width=5.0cm]{fig12a.eps}  
 \includegraphics[width=3.4cm]{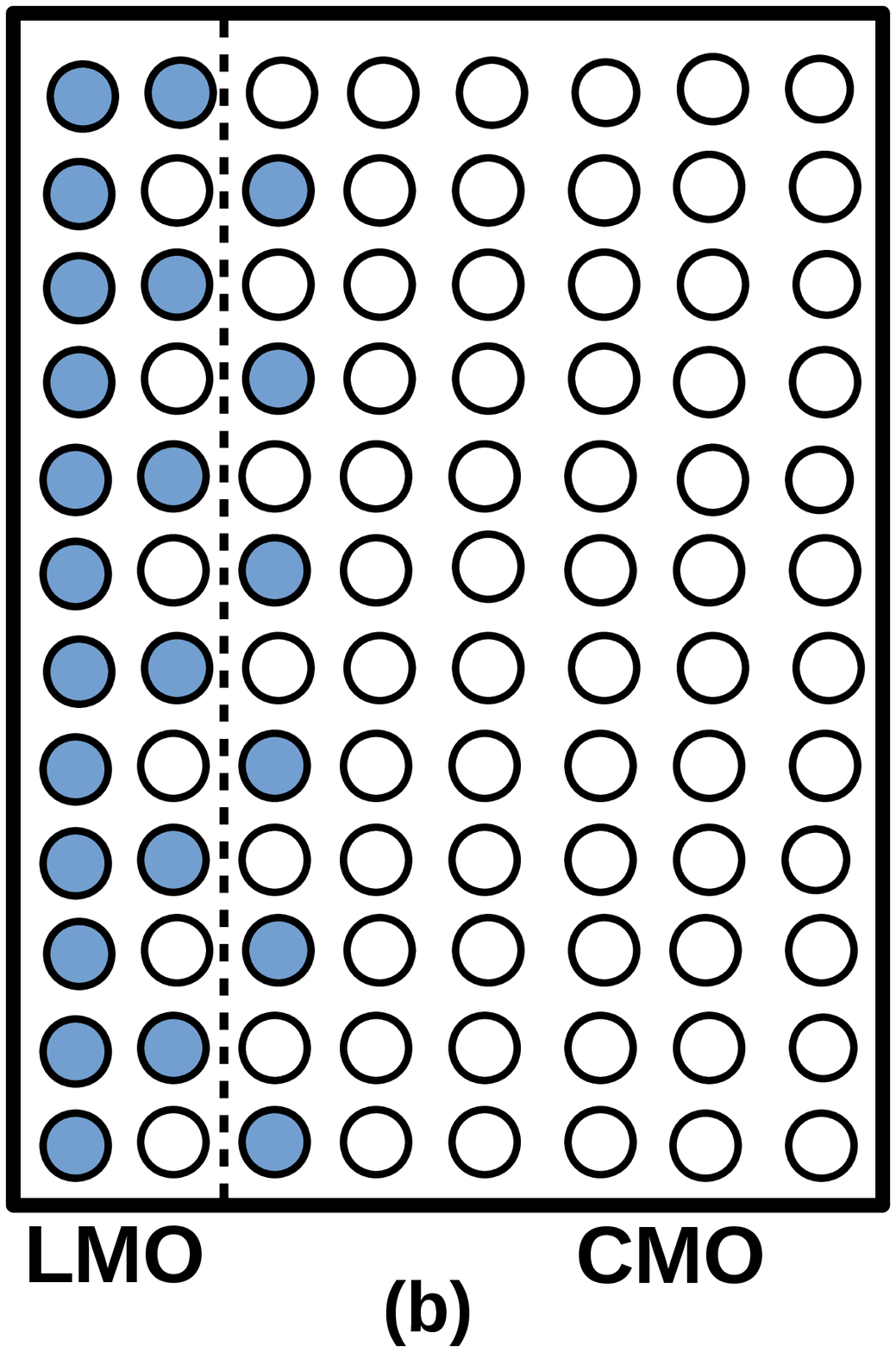} \\
 \vspace{0.5cm}
 \includegraphics[width=5.0cm]{fig12c.eps}  
 \includegraphics[width=3.4cm]{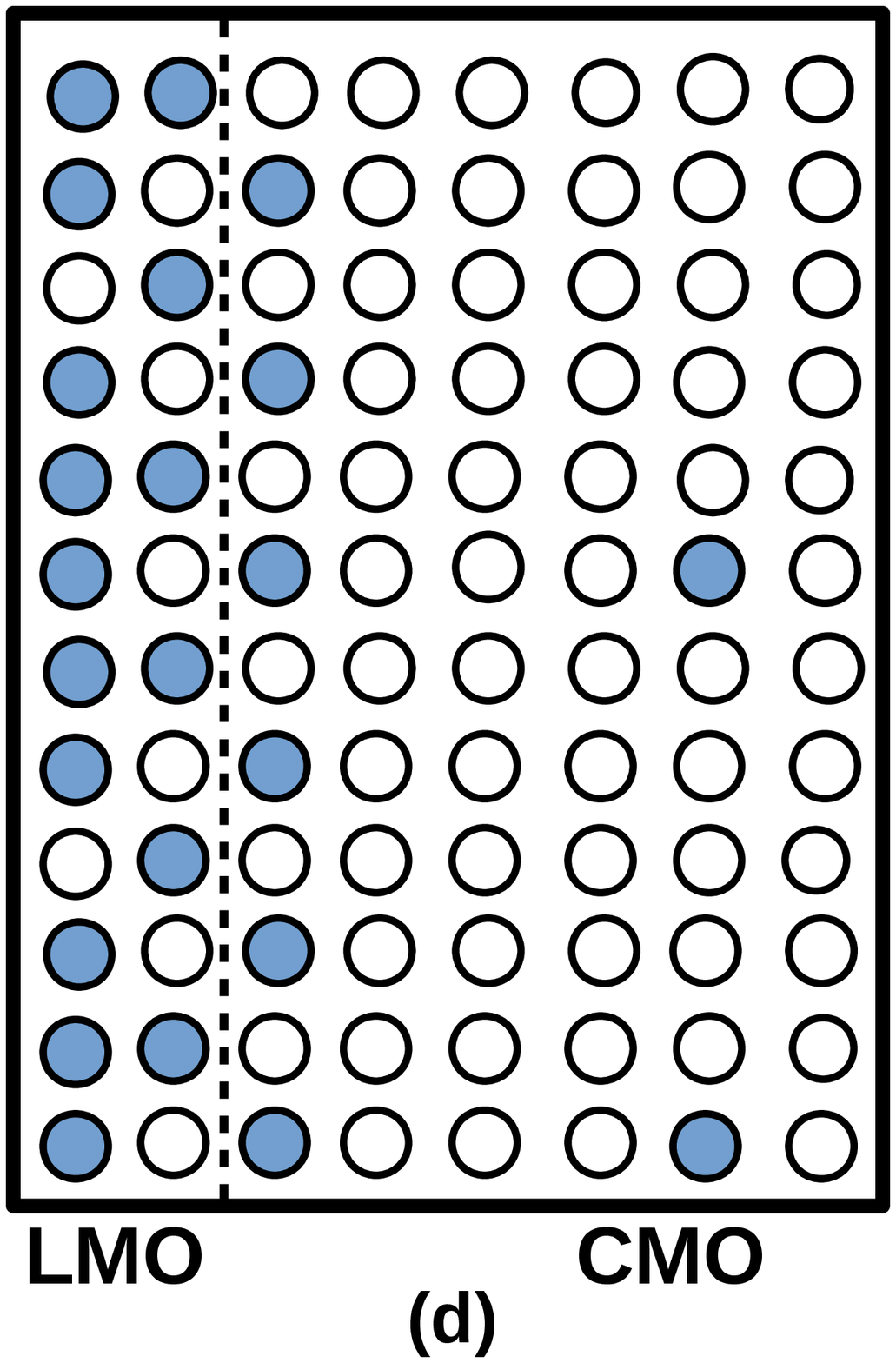} \\
\caption{(Color online)
In asymmetric heterostructure defined on a $12\times8$ lattice with 2 layers of LMO 
and 6 layers of CMO, when coupling $g=2.0$, 
(a) at ${\rm E_{ext}} = 0$, layer-averaged charge density $\langle n(I) \rangle $ and
layer-averaged magnetization $\langle m(I) \rangle $ of normalized-to-unity $t_{\rm 2g}$ spins; 
(b) at ${\rm E_{ext}} = 0$, ground state configuration;
(c) at ${\rm E_{ext}} = 300 ~{\rm kV/cm}$, layer-averaged charge profile 
$\langle n(I) \rangle $ and layer-averaged magnetization
profile $\langle m(I) \rangle $ of normalized-to-unity $t_{\rm 2g}$ spins; and
(d) at ${\rm E_{ext}} = 300 ~{\rm kV/cm}$, ground state configuration.}
 \label{asym_lmo2_smo6_e0}
  \end{figure}

  We have an asymmetry here regarding the number of layers;  
  the LMO side has two layers whereas the CMO side has six layers.
  For zero external electric field, as shown in Figs. \ref{asym_lmo2_smo6_e0}(a) and \ref{asym_lmo2_smo6_e0}(b),
the two interfacial layers 2 and 3
contain perfectly alternate arrangement of electrons and holes; hence, the layers are fully polarized. 
Layer 1 is also ferromagnetically aligned with layer 2 because of their proximity.
Beyond layer 3, similar to the G-AFM order in bulk CMO, the CMO side  is antiferromagnetic, 
resulting in zero magnetization.
Turning on the sizeable electric field ${\rm E_{ext}} = 300 ~{\rm kV/cm}$,
as depicted in Figs. \ref{asym_lmo2_smo6_e0}(c) and \ref{asym_lmo2_smo6_e0}(d),
leads to a few electrons
from layer 1 ending up in a farther layer 7; resultantly, the magnetic polarons in layer 7
partially polarize it. It is interesting to note that, here too charge modulation occurs
due to the Coulomb term $H_{\rm coul}$  as demonstrated through Eq. (\ref{E_coul}).
There is an overall increase
 in the magnetization 
 (of normalized-to-unity $t_{\rm 2g}$ spins)
 by 0.04/site; thus, the magnetoelectric effect is not huge.
\subsubsection{Asymmetric $12\times8$ system of 6  LMO layers and 2 CMO layers with $g=2.0$ 
and ${\rm E_{ext}} = 300 ~{\rm kV/cm}$}
\label{asym_6_2_12x8_g2_e300}
  \begin{figure}[t]
 \includegraphics[width=5.0cm]{fig13a.eps} 
 \includegraphics[width=3.4cm]{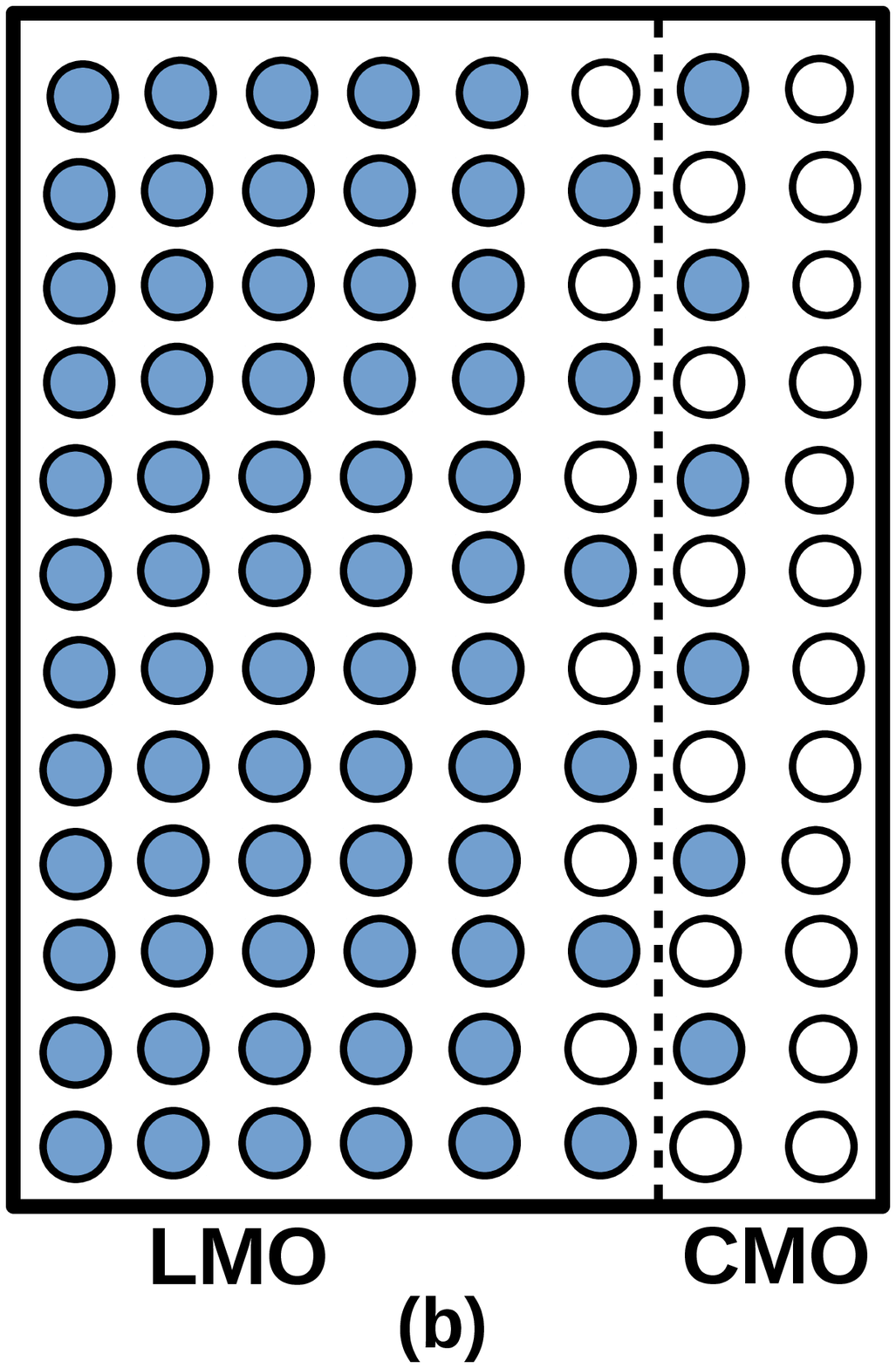} \\
 \vspace{0.5cm}
 \includegraphics[width=5.0cm]{fig13c.eps} 
 \includegraphics[width=3.4cm]{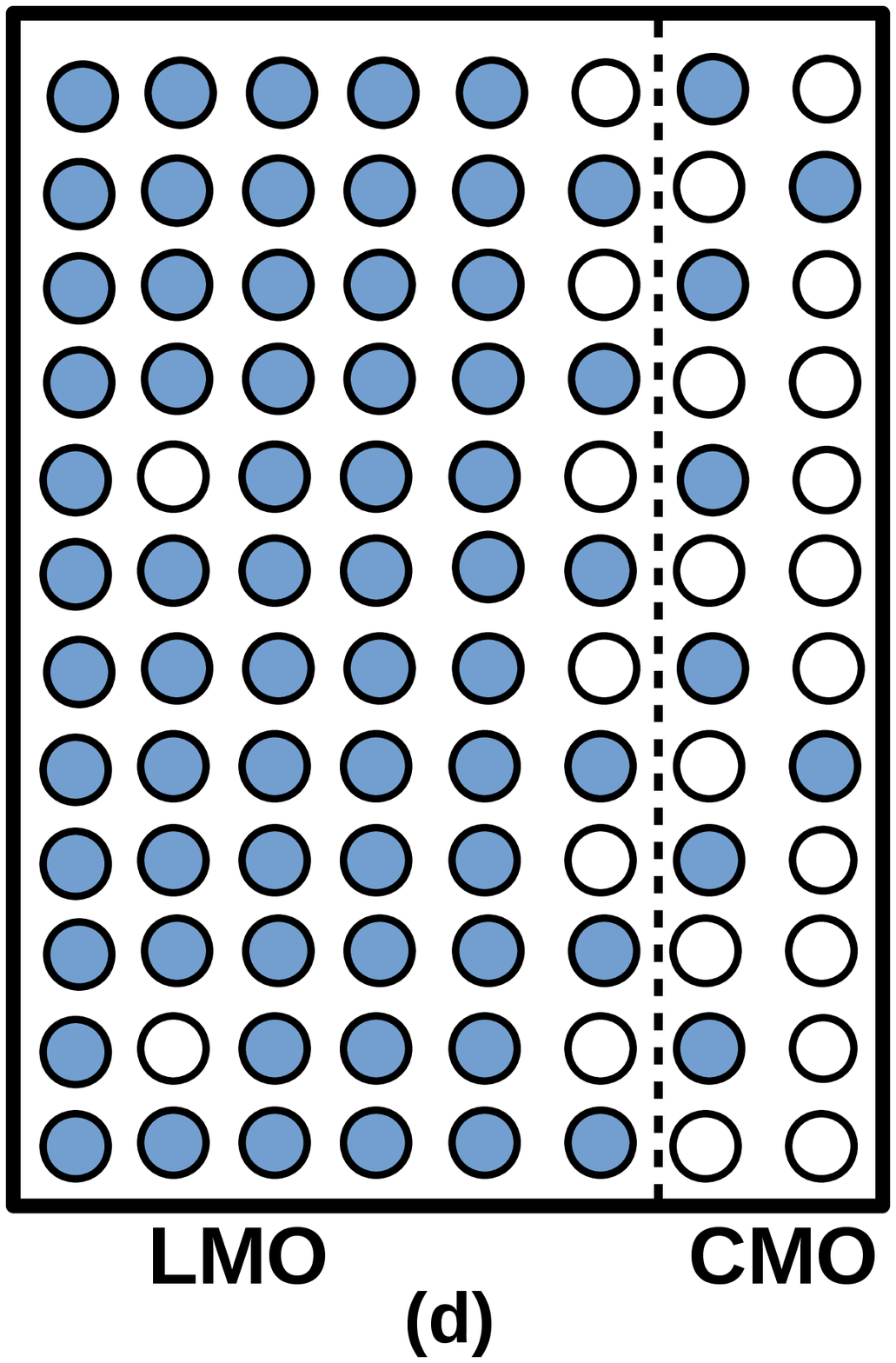} \\
\caption{(Color online)
In asymmetric $12\times8$ lattice with 6 layers of LMO 
and 2 layers of CMO, when electron-phonon interaction $g=2.0$,
(a) at external electric field ${\rm E_{ext}} = 0$, layer-averaged electron density $\langle n(I) \rangle $ and
layer-averaged magnetization $\langle m(I) \rangle $ of $t_{\rm 2g}$ spins normalized to unity; 
(b) at ${\rm E_{ext}} = 0$, ground state electronic configuration;
(c) at ${\rm E_{ext}} = 300 ~{\rm kV/cm}$, layer-averaged electron density 
$\langle n(I) \rangle $ and layer-averaged magnetization
 $\langle m(I) \rangle $ of $t_{\rm 2g}$ spins  normalized to unity; and
(d) at ${\rm E_{ext}} = 300 ~{\rm kV/cm}$, ground state.}
 \label{asym_lmo6_smo2_e0}
  \end{figure}

  Here, compared to the previous structure in Fig. \ref{asym_lmo2_smo6_e0},
   we  have the 
opposite asymmetric structure of 6 LMO layers and 2 CMO layers (see Fig. \ref{asym_lmo6_smo2_e0}).
Due to the
asymmetry connection between these two structures, we obtain a mirror image of the previous
charge configuration for both with and without the external electric field.
 The interfacial layers 6 and 7, as expected, are totally ferromagnetic.
 Furthermore, layer 5 is also ferromagnetically aligned with layers 6 and 7
  due to the ferromagnetic couplings $J_{\rm eh}$ and $J_{\rm xy}$.
 At zero external field [as shown in Figs. \ref{asym_lmo6_smo2_e0}(a) and \ref{asym_lmo6_smo2_e0}(b)],
 similar to the bulk situation,
 we have A-AFM on the LMO side up to layer 5;
 layer 8, similar to the bulk CMO, is also antiferromagnetic.
 The minority carriers, generated due to the external electric field ${\rm E_{ext}} = 300 ~{\rm kV/cm}$
 [as shown in Figs. \ref{asym_lmo6_smo2_e0}(c) and \ref{asym_lmo6_smo2_e0}(d)],
  produce magnetic polarons which increase the polarization.
 There 
 is an overall increase in  the magnetization 
 (of normalized-to-unity $t_{\rm 2g}$ spins)
 by about 0.1/site implying
 a large magnetoelectric effect.

 From the various symmetric and asymmetric LMO-CMO configurations considered, we
 conclude that the symmetric arrangement yields the largest magnetoelectric effect.\\

\section{CONCLUSIONS}
We used the heuristic notion that, when the two Mott insulators LMO and CMO are brought together to form
a heterostructure, one may realize the entire phase diagram of ${\rm La_{1-x}Ca_x MnO_3}$ 
across the heterostructure with the $x=0$
phase occurring at the LMO surface at one end and 
 evolving to the $x=0.5$ state at the oxide-oxide
 interface and finally to the  $x=1$ phase at the other end of CMO. 
 However, owing to the reduced dimensions (namely, quasi two-dimensions),
 we expect only A-AFM and FMI
  phases on the LMO side. 
 On enhancing the minority carrier density (on both sides of the interface)
 by using a sizeable external electric field, we showed that the FMI region can be further expanded
at the expense of the A-type AFM region on the LMO side and the G-AFM domain on the CMO side,
thereby producing a giant magnetoelectric
effect.
{It is important to note that the system behaves  like an anisotropic Coulombic solid:
for a given average density in a layer, the minority charges by and large order periodically
and as far apart as possible;
the Coulombic interaction dictates how the charge arrangement of one layer adjusts itself 
with respect to the configuration in another layer.  
Furthermore,  the two layers  at the interface are ferromagnetically ordered as they are at half-filling.
When
electric fields are introduced in the system, the minority charge density away from the interface
gets enhanced
leading to charge reordering; consequently, the  magnetization away from the interface changes layer by layer
resulting in an  increase in the total ferromagnetic moment.
This scenario is key to the understanding of the magnetoelectric effect. }

As a guide to designing magnetoelectric devices,
we find that symmetric heterostructures with equal number of LMO and CMO layers
yield larger magnetoelectric effect compared to asymmetric heterostructures
with unequal number of LMO and CMO layers. 
{
It should also be noted that this heterostructure/device
can be used 
near helium liquefaction temperatures; 
on the other hand,
the  magnetoelectric function disappears before  nitrogen liquefaction temperature is attained.}

 We also would like to mention that if a superlattice were formed from the heterostructure
${\rm (Insulator)/(LaMnO_3)_n/(CaMnO_3)_n/(Insulator)}$, then the dipoles from each repeating
heterostructure unit will add up to produce a giant electric dipole moment;
furthermore, this superlattice will also realize a giant magnetoelectric effect.
 On the other hand, if the insulator
layers were not present, the superlattice formed from the repeating unit
${\rm (LaMnO_3)_n/(CaMnO_3)_m}$ 
would not produce a large electric dipole as the charge from
 LMO can leak to CMO on both sides leading to a small net dipole moment.
 More importantly, the 
${\rm (LaMnO_3)_n/(CaMnO_3)_m}$ superlattice will also not generate a giant magnetoelectric effect
as an applied electric field will not alter much the total amount of charge in LMO or CMO;
this is because charge leaked by LMO to CMO on one side is replaced by CMO on the other side.

 {Based on the above arguments, it should be clear that,
 compared to experiments involving superlattices where alloy/bulk effect vanishes
 when the thinner  side of the repeating unit has more than two  layers,
 in our heterostructure bulk nature should vanish when the thinner side has more than 1 layer. Thus, in 
 the superlattice ${\rm (LaMnO_3)_{2n}/(SrMnO_3)_n}$ studied in Ref. \onlinecite{anand1},
 the metallic behavior (corresponding to bulk ${\rm La_{0.67}Sr_{0.33}MnO_3}$) disappears
 for $n> 2$ and is replaced by insulating behavior; whereas, in the heterostructure 
  ${\rm (Insulator)/(LaMnO_3)_{2n}/(SrMnO_3)_n/(Insulator)}$ we expect insulating
  behavior for $n > 1$. 
    {This observation supports our assumption that,
    in our quasi 2D LMO-CMO heterostructures 
{ {(corresponding to
  LCMO that has a narrower band width and a stronger electron-phonon coupling compared to LSMO)}, 
   only a single narrow-width polaronic band is pertinent. 
 Additionally, 
  interfacial roughness (if considered) will further
 reduce the band width and suppress metallicity.}}

Lastly, it should also be pointed out that, in a realistic situation, we have electron-electron repulsion 
 (produced by cooperative electron-phonon interaction)
 and double-exchange generated ferromagnetic coupling extending to next-nearest-neighbor sites \cite{ravindra2}
  leading to a larger
  magnetic polaron and thus producing a stronger magnetoelectric effect compared to what
  our calculations reveal.

\section{acknowledgements} 
S. Y. acknowledges discussions with T. V. Ramakrishnan, D. D. Sarma, and K. Pradhan.
S. P. acknowledges  discussions with S. Nag, R. Ghosh, S. Kadge, N. Swain,  S. Mukherjee, and R. Raman. 
Work at Argonne National
Laboratory is supported by the U.S. Department of
Energy, Basic Energy Sciences Materials Science and Engineering, under contract
no. DE-AC02-06CH11357.

\end{document}